%% file: ttx_qcd.tex
\documentclass[a4paper,11pt]{article}
\usepackage{jheppub} 
\usepackage[T1]{fontenc}
\usepackage{booktabs}
\usepackage[dvips,table]{xcolor} 
\usepackage{xspace}
\usepackage{dcolumn}
\usepackage{hyperref}
\usepackage{caption}
\usepackage
[subrefformat=parens,position=top,skip=-15pt,margin=15pt,justification=justified,singlelinecheck=false]
{subcaption}

\newcommand{\eq}[1]{\begin{equation} #1 \end{equation}}

\newcommand{\changed}[1]{#1}

\allowdisplaybreaks[1]

\input{macros}

\title{Off-shell production of top--antitop pairs in the lepton+jets channel at NLO QCD}
\subheader{\today}

\author{Ansgar Denner,}
\author{Mathieu Pellen}

\affiliation{
        Universit\"at W\"urzburg, %
        Institut f\"ur Theoretische Physik und Astrophysik, \\  %
        97074 W\"urzburg, %
        Germany%
}

\emailAdd{ansgar.denner@physik.uni-wuerzburg.de}
\emailAdd{mathieu.pellen@physik.uni-wuerzburg.de}

\abstract{The production of top-quark pairs that subsequently decay
  hadronically and leptonically (lepton+jets channel) is one of the
  key processes for the study of top-quark properties at the LHC.  In
  this article, NLO QCD corrections of order
  $\order{\alphas^3\alpha^{4}}$ to the hadronic process $\Pp\Pp\to
  \mu^-\bar{\nu}_\mu\Pb\bar{\Pb} \Pj \Pj$ are presented.  The
  computation includes off-shell as well as non-resonant
  contributions, and experimental event selections are used in order
  to provide realistic predictions.  The results are provided in the
  form of cross sections and differential distributions.  The
  QCD corrections are sizeable and different from the ones of the fully
  leptonic channel.  This is due to the different final state where
  here four jets are present at leading order.
  }
\begin{document}

\maketitle
\flushbottom

\newpage

\section{Introduction}

The large amount of data collected during the run~II of the Large
Hadron Collider (LHC) will allow to probe the high-energy behaviour of
many Standard Model processes.  This is particularly important as the
high-energy tails of differential distributions are expected to be
sensitive to new-physics contributions.  In order to achieve stringent
tests of the Standard Model, theoretical predictions should be
computed with at least next-to-leading order (NLO) QCD and electroweak
(EW) accuracy.  In addition, in order to be directly comparable with
experiments, the calculations should be differential in the final
states that are actually measured in experiments.  The last point is
particularly crucial as in the high-energy tails of differential
distributions, off-shell and non-resonant contributions become
increasingly relevant.  Thus, theoretical predictions should include
as much as possible off-shell as well as non-resonant effects in order
to describe appropriately the final states seen experimentally.

In this regard, the production of top--antitop pairs is exemplary.
In the past few years, several off-shell computations have been performed for this process.
First, NLO QCD corrections
\cite{Denner:2010jp,Bevilacqua:2010qb,Denner:2012yc,Frederix:2013gra,Cascioli:2013wga} have been
calculated and matched to parton shower in the narrow-width approximation \cite{Campbell:2014kua} and recently accounting
for the resonance structure of the process \cite{Jezo:2016ujg}.  The NLO EW corrections
have been computed recently \cite{Denner:2016jyo}.  In addition, the
results for the off-shell production of a top--antitop pair in
association with a jet at NLO QCD
\cite{Bevilacqua:2015qha,Bevilacqua:2016jfk} or in association with a
Higgs boson at NLO QCD and EW \cite{Denner:2015yca,Denner:2016wet} are
available.  Recently, an approximate NNLO QCD computation including
decays \cite{Gao:2017goi} was published.  This seems to
reproduce well the full NNLO QCD results
\cite{Czakon:2016ckf,Czakon:2016dgf} for on-shell top-quark production
but does not account for non-resonant top-quark contributions that can
be significant
\cite{Heinrich:2013qaa,Heinrich:2017bqp,Bevilacqua:2017ipv}.

For now, all these computations have focused on the channel where both
top quarks decay leptonically.  From the theoretical point of view,
this channel is preferred as it contains only two strongly interacting particles in the
final state (two bottom quarks).  However, experimentally, the channel
where one top quark decays hadronically (denoted hadronic top quark)
while the other decays leptonically (denoted leptonic top quark) is
also investigated \cite{Aaboud:2017fha,Khachatryan:2016mnb}.  It is
dubbed lepton+jets channel, features a larger cross section, and has
the advantage to allow for a better reconstruction of the event as
only one neutrino contributes to the missing transverse energy (as
opposed to the fully leptonic channel where two neutrinos carry away
some momentum).

For this reason, we have computed for the first time the NLO QCD
corrections to the production of top--antitop pairs in the lepton+jets
channel, \emph{i.e.}\ the process $\Pp\Pp\to
\mu^-\bar{\nu}_\mu\Pb\bar{\Pb} \Pj \Pj$.  We have considered the order
$\order{\alphas^2\alpha^{4}}$ contributions at leading order (LO) and present the NLO
QCD corrections at the order $\order{\alphas^3\alpha^{4}}$.  The
computation features all off-shell and non-resonant effects to the
partonic channels that involve two resonant top quarks.
In
particular, it allows for a direct comparison with experimental
measurements as the event selection applied to the final state follows
the experimental one.  The corrections are sizeable and different from
the ones to the top-pair-production process with two leptonically
decaying top quarks.  In particular, they can be much larger in some
phase-space regions.  This originates from the different final state
where here four jets are present at LO (two light jets and two bottom
jets).  In particular, the increased number of jets in the final state
and the corresponding irreducible background  can alter the
predictions significantly.  
This is discussed in detail at the level of the fiducial cross section
and in several differential distributions.  More precisely, new
effects show up in the tails of the transverse momentum distributions
as well as in other regions that are sizeably affected by non-resonant
contributions.

\changed{In this article we are focusing on the NLO QCD corrections to
  the production of a pair of top quarks and the corresponding
  off-shell effects. We do not include suppressed contributions such
  as partonic channels that do not involve two resonant top quarks and
  interferences of amplitudes of order $\order{g_\mathrm{s}^4 g^{2}}$
  with those of order $\order{g^{6}}$. Furthermore, we do not take
  into account bottom-quark-induced and photon-induced contributions,
  which are suppressed owing to the involved parton distributions
  functions (PDFs). The corresponding LO contributions are at the
  per-mille level of the fiducial cross section for top-pair
  production and thus negligible with respect to the experimental
  precision at the LHC.  Since all these suppressed contributions are
  of the order of the numerical accuracy of our NLO predictions and
  not visible in the presented results for distributions, we decided
  not to include them.}

The article is organised as follows: in \refse{sec:definition} the
process studied is defined, while in \refse{sec:details} the
technical details of the calculation are presented.  Section
\ref{sec:results} is devoted to the numerical results and their
discussion.  In particular, cross sections as well as differential
distributions are presented.  Finally, \refse{sec:conclusion}
contains a summary and concluding remarks.

\section{Definition of the process}
\label{sec:definition}

\begin{figure}
\newcommand{\myframebox}{\framebox}
\renewcommand{\myframebox}{\relax}
  \begin{subfigure}{1\linewidth}
                \myframebox{
                        \includegraphics[width=\linewidth]{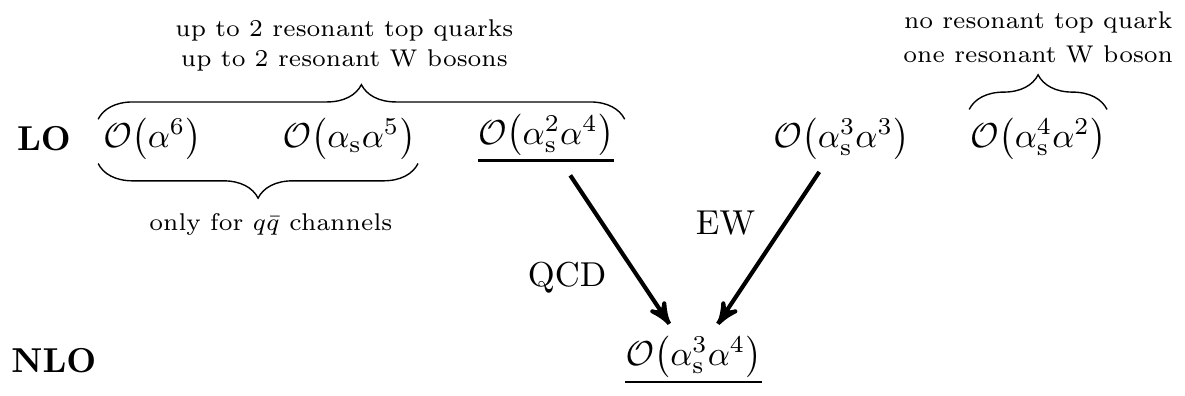}
                } \vspace*{-2ex}
  \end{subfigure}
  \caption{Graphical representation of contributions to the cross
    section of $\Pp\Pp\to \mu^-\bar{\nu}_\mu\Pb\bar{\Pb} \Pj\Pj$.  At
    NLO, the order $\order{\alphas^3\alpha^{4}}$ receives QCD
    corrections and EW correction to the orders
    $\order{\alphas^2\alpha^{4}}$ and $\order{\alphas^3\alpha^{3}}$,
    respectively.  The two underlined contributions
    [$\order{\alphas^2\alpha^{4}}$ at LO and
    $\order{\alphas^3\alpha^{4}}$ at NLO] are the ones considered in
    the present calculation.}
\label{fig:allorder}
\end{figure}

\begin{figure}
  \newcommand{\myframebox}{\framebox}
  \renewcommand{\myframebox}{\relax}
  \begin{subfigure}{0.32\linewidth}
    \myframebox{
      \includegraphics[width=\linewidth]{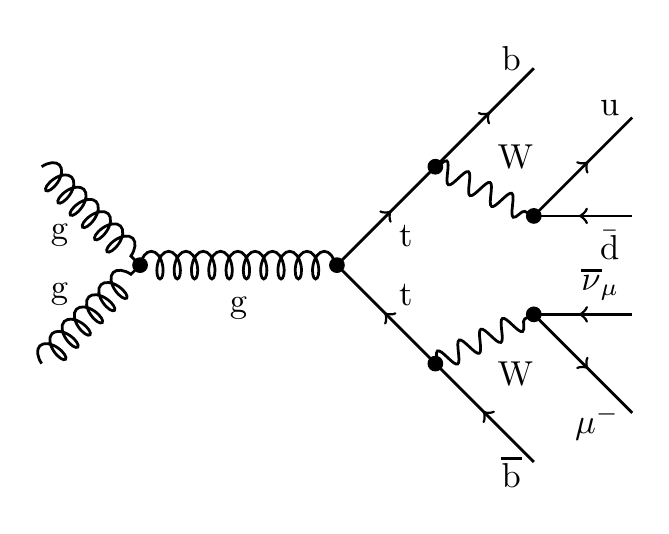}
      }
   \end{subfigure}
   \begin{subfigure}{0.34\linewidth}
     \myframebox{
       \includegraphics[width=\linewidth]{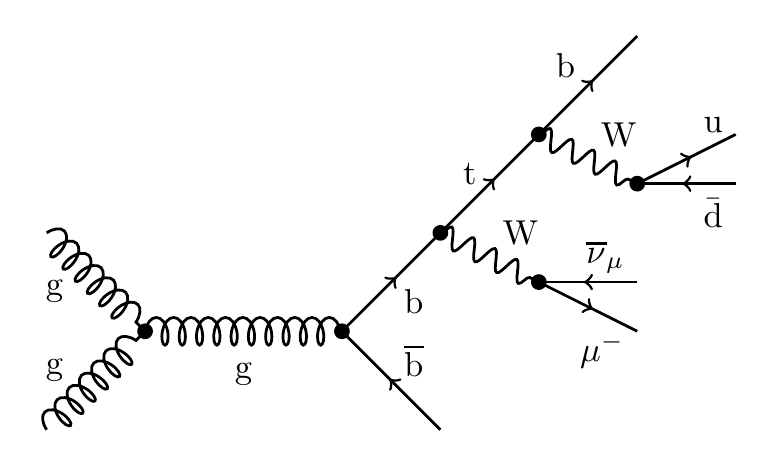}
       }
   \end{subfigure}
   \begin{subfigure}{0.32\linewidth}
     \myframebox{
       \includegraphics[width=\linewidth]{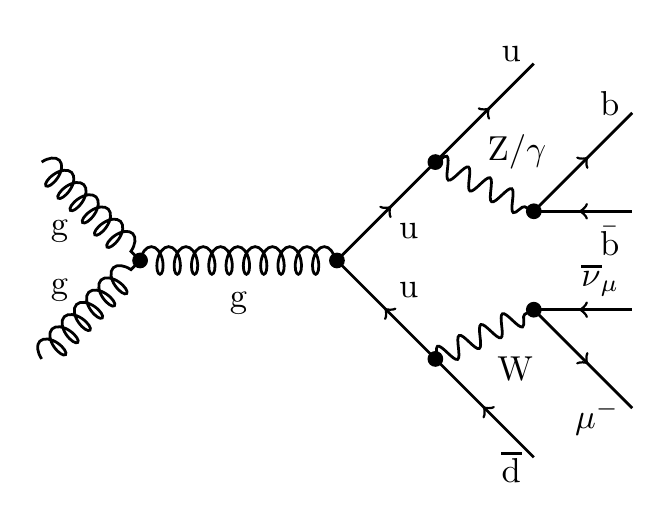}
       } 
   \end{subfigure} \vspace*{-2ex}
   \caption{Sample tree-level Feynman diagrams of order
     $\order{g_{\rm s}^2g^{4}}$ for $\Pg\Pg\to \mu^-\bar{\nu}_\mu\Pb\bar{\Pb} \Pj\Pj$.
     Some diagrams have two resonant top quarks and two resonant W
     bosons (left) while some have only one resonant top quark and two
     resonant W~bosons (middle) or no resonant top quark and one
     resonant W~boson (right).}
        \label{fig:LO}
\end{figure}

We consider the off-shell production of top--antitop pairs in the
lepton+jets channel at the LHC, \emph{i.e.}\ the hadronic process
\eq{\Pp\Pp\to \mu^-\bar{\nu}_\mu\Pb\bar{\Pb} \Pj \Pj. }
At the matrix-element level, this process possesses three types of LO contributions of orders  $\order{g^{6}}$,
$\order{g_{\rm s}^2g^{4}}$, and $\order{g_{\rm s}^4g^{2}}$.
The corresponding contributions at the cross-section level are 
shown in \reffi{fig:allorder}.
Among these, the dominant one is of order
$\order{\alphas^2\alpha^{4}}$.  Sample diagrams contributing at the
order $\order{g_{\rm s}^2g^{4}}$ are displayed in \reffi{fig:LO}.
There are contributions involving two resonant top quarks and two
resonant W~bosons (left), contributions with one resonant top quark
and two resonant W~bosons (middle), and contributions with no resonant
top quark and one resonant W~boson (right).
The two contributions of orders $\order{\alpha^{6}}$ and
$\order{\alphas\alpha^{5}}$ are suppressed owing to power counting in
the two coupling constants and because they exist only for $q \bar q$
channels (which is suppressed with respect to the gg channel at the
LHC).  The contributions of orders $\order{\alphas^3\alpha^{3}}$ and
$\order{\alphas^4\alpha^{2}}$ are suppressed due to the absence of
doubly resonant top quarks/W~bosons.  The NLO QCD corrections to the
dominant contribution are thus of order $\order{\alphas^3\alpha^{4}}$.
As described in \reffi{fig:allorder}, these NLO corrections consist of
QCD and EW corrections to the orders $\order{\alphas^2\alpha^{4}}$ and
$\order{\alphas^3\alpha^{3}}$, respectively.  \changed{Contributions
  at order $\order{\alphas^2\alpha^{4}}$ also arise from the
  interference of amplitudes of orders $\order{g_{\rm s}^4g^{2}}$ and
  $\order{g^{6}}$.  These contributions are strongly suppressed, since
  they only arise in $q \bar q$ channels and the $\order{g_{\rm
      s}^4g^{2}}$ amplitude does not involve resonant top quarks and
  only one resonant W boson.  The corresponding LO contributions are
  at the level of $10^{-6}$ for the fiducial cross section and
  therefore completely negligible.}

Using symmetries between different quark families, the fully leptonic
process can be built from only four independent partonic processes:
the ones with initial states $\Pg\Pg$, $\Pu\bar{\Pu}/\bar{\Pu}\Pu$,
$\Pd\bar{\Pd}/\bar{\Pd}\Pd$, and $\Pb\bar{\Pb}/\bar{\Pb}\Pb$.  For the
semi-hadronic decay of the top-quark pair, the number of independent
partonic channels rises to 32. 
\changed{Among  these, the six partonic
channels that feature two resonant top quarks approximate the LO fiducial
cross section at the level of per mille ($0.28\%$) for the set-up described
in \refses{subsec:input} and \ref{subsec:cuts}.
The other channels, which can be constructed upon crossing one or two
final state quarks in the initial state involve at most one top-quark
and one W-boson resonance. They are further suppressed by a
di-jet invariant-mass cut. Relaxing this cut, these contributions
become of the order of a couple of per cent ($2.0\%$) of the fiducial
cross section at LO.}
The channels involving two resonant top quarks read:
\begin{align}
\label{eq:bornprocesses}
 \Pg\Pg &\to \mu^-\bar{\nu}_\mu\Pb\bar{\Pb} q_i \bar q_j, & q_i q_j \in \{ \Pu \Pd, \Pc \Ps \}, \nonumber \\
 q_i \bar q_i/ \bar q_i q_i &\to \mu^-\bar{\nu}_\mu\Pb\bar{\Pb} q_i \bar q_j, & q_i q_j \in \{ \Pu \Pd, \Pc \Ps \}, \nonumber \\
 q_i \bar q_i/ \bar q_i q_i &\to \mu^-\bar{\nu}_\mu\Pb\bar{\Pb} q_j \bar q_k, & q_i q_j q_k \in \{ \Pu \Pc \Ps, \Pc \Pu \Pd \}, \nonumber \\
 q_i \bar q_i/ \bar q_i q_i &\to \mu^-\bar{\nu}_\mu\Pb\bar{\Pb} q_j \bar q_i, & q_i q_j \in \{ \Pd \Pu, \Ps \Pc \}, \nonumber \\
 q_i \bar q_i/ \bar q_i q_i &\to \mu^-\bar{\nu}_\mu\Pb\bar{\Pb} q_j \bar q_k, & q_i q_j q_k \in \{ \Pd \Pc \Ps, \Ps \Pu \Pd \}, \nonumber \\
 \Pb \bar \Pb / \bar \Pb \Pb &\to \mu^-\bar{\nu}_\mu\Pb\bar{\Pb} q_i \bar q_j, & q_i q_j \in \{ \Pu \Pd, \Pc \Ps \}.
\end{align}%
\changed{For the fiducial cross section considered in this work,
the LO bottom-quark contributions turn out to be $0.13\%$. Since this
is below the integration error of the NLO calculation ($0.5\%$) we do
not include them  in the cross sections and differential distributions
presented in this article.}
Therefore, in the following computation, only the five remaining
partonic channels and the corresponding NLO QCD corrections are
considered.

\changed{In addition there are contributions from photon-induced
  channels, which are, however, suppressed by the photon PDFs.  The
  leading photon-induced contribution arises from the process $\Pg
  \gamma / \gamma \Pg \to \mu^-\bar{\nu}_\mu\Pb\bar{\Pb} q_l \bar q_m$
  at order $\order{\alphas\alpha^{5}}$. It is enhanced owing to the
  gluon PDF and the fact that it possesses doubly-resonant top-quark
  contributions, but it is suppressed by a factor $\alpha/\alphas$
  with respect to the leading $\Pg\Pg$- and $q\bar{q}$-induced
  contributions. At the order $\order{\alphas\alpha^{5}}$, it amounts
  to $0.31\%$ of the LO fiducial cross section.  These findings are in
  line with the ones of \citeres{Denner:2016jyo,Denner:2016wet} given
  that the present number has been obtained with the
  $\mathrm{LUXqed}\_\mathrm{plus}\_\mathrm{PDF4LHC15}\_\mathrm{nnlo}\_100$
  set~\cite{Manohar:2016nzj}.  All other photon-induced contributions
  at LO or NLO can only be a fraction of these because of PDF
  suppression, coupling suppression, and/or lacking resonance
  enhancement.}

\section{Details of the calculation}
\label{sec:details}

The Monte Carlo program used for this computation has already been
employed for NLO computations involving off-shell top quarks
\cite{Denner:2015yca,Denner:2016jyo,Denner:2016wet}.  In addition,
this program has also been used for the NLO QCD and EW computation of
the process $\Pp \Pp \to \mu^+\nu_\mu \Pe^+ \nu_{\Pe} \Pj \Pj$
\cite{Biedermann:2017bss} where also two QCD jets are present at LO.
Finally, the program uses similar phase-space mappings to those of
\citeres{Berends:1994pv,Denner:1999gp,Dittmaier:2002ap} to ensure a
fast integration even for processes with high multiplicities.

\paragraph{Virtual corrections:}

\begin{figure}
  \newcommand{\myframebox}{\framebox}
  \renewcommand{\myframebox}{\relax}
  \begin{subfigure}{0.32\linewidth}
    \myframebox{
      \includegraphics[width=\linewidth]{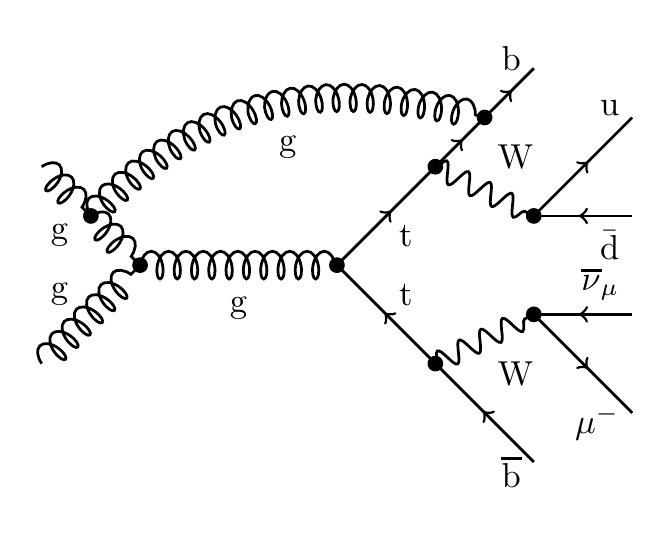}
      }
  \end{subfigure}
  \begin{subfigure}{0.32\linewidth}
    \myframebox{
      \includegraphics[width=\linewidth]{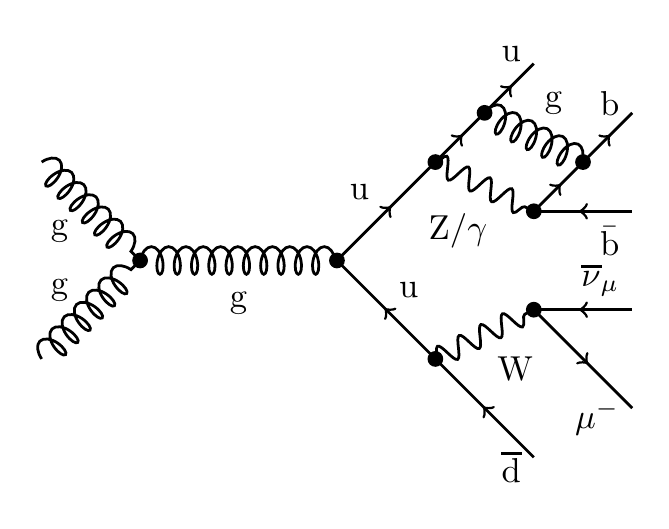}
      }
  \end{subfigure}
  \begin{subfigure}{0.32\linewidth}
    \myframebox{
      \includegraphics[width=\linewidth]{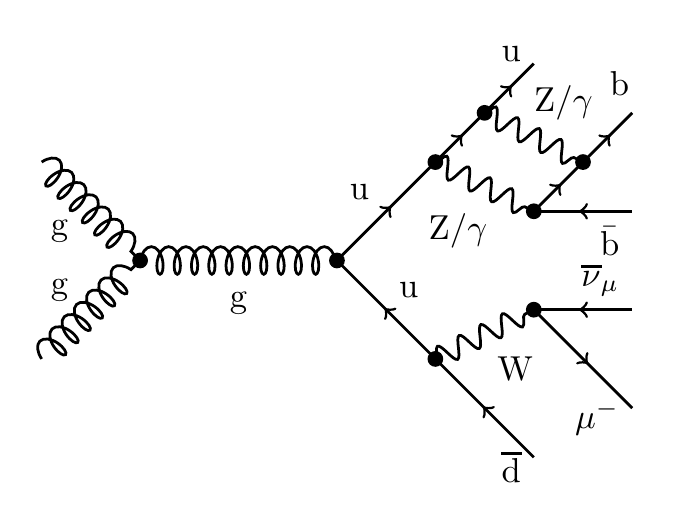}
      }
  \end{subfigure} \vspace*{-2ex}
  \caption{Sample one-loop Feynman diagrams contributing to $\Pg\Pg\to
    \mu^-\bar{\nu}_\mu\Pb\bar{\Pb} \Pj\Pj$ at order
    $\order{\alphas^3\alpha^{4}}$ at NLO.  While some diagrams can be
    uniquely identified as
    QCD corrections (left) or  EW corrections (right), this is not
    possible for others (middle).}
        \label{fig:NLO}
\end{figure}

We include virtual corrections obtained from all one-loop amplitudes
interfered with tree amplitudes giving rise to an order
$\order{\alphas^3\alpha^{4}}$ contribution for all the processes described in
Eq.~\eqref{eq:bornprocesses}. This includes, in particular, one-loop
amplitudes of order $\order{g_s^4 g^4}$ interfered with $\order{g_s^2
  g^4}$ tree amplitudes.  Such loop amplitudes are obtained upon
inserting a gluon into the tree-level diagrams of order $\order{g_s^2 g^4}$
for the processes of Eq.~\eqref{eq:bornprocesses}.  Examples are shown
in the left and the middle of \reffi{fig:NLO}.

Virtual corrections of order $\order{\alphas^3\alpha^{4}}$ can also
been obtained by interfering one-loop amplitudes of order
$\order{g_s^2 g^6}$ [which would usually be referred to as EW
corrections to the dominating LO diagrams of order $\order{g_s^2
  g^4}$] with $\order{g_s^4 g^2}$ tree-level amplitudes.  Such an EW
one-loop diagram is depicted on the right-hand side of
\reffi{fig:NLO}, while a tree-level diagram of order $\order{g_s^4
  g^2}$ would for instance result from the one in the right-hand side
of \reffi{fig:LO} upon replacing the Z boson or photon by a gluon. LO
contributions of order $\order{g_s^4g^{2}}$ exist only for
the process $\Pp\Pp\to \mu^-\bar{\nu}_\mu\Pb\bar{\Pb} \Pj \Pj$ but not
for the $\mu^-\bar{\nu}_\mu\Pb\bar{\Pb} \Pe^+\nu_\Pe$ final state
relevant for two leptonically decaying top quarks.  The situation is
similar to the case of NLO EW corrections to the fully leptonic
process where interferences with one-loop QCD corrections have to be
considered \cite{Denner:2016jyo}.  We note that these contributions
are numerically small as they feature only one resonant W~boson but
they must be included in order to ensure an infrared (IR) finite
result.

Finally virtual corrections of order $\order{\alphas^3\alpha^{4}}$
also result upon interfering one-loop amplitudes of order
$\order{g_s^6 g^2}$, \emph{i.e.}\ QCD corrections to the suppressed LO
diagrams of order $\order{g_s^4 g^2}$, with $\order{g^6}$ tree-level
amplitudes. These corrections can be separated on the basis of Feynman
diagrams and have been neglected.
Since such contributions can be uniquely identified as QCD corrections, the associated IR singularities cancel upon adding the corresponding real-radiation contributions.
Numerically, they are well below the per-mille level of the LO cross section and are thus irrelevant.

All the tree-level and one-loop matrix elements have been obtained
from the public code \recola
\cite{Actis:2012qn,Actis:2016mpe}.\footnote{Note that \recola has been
  recently implemented in the multi-purpose Monte Carlo codes {\sc
    Sherpa}~\cite{Biedermann:2017yoi} and {\sc Whizard}
  \cite{whizard:2017,Kilian:2018onl} which allows to compute NLO QCD and EW
  corrections.}
It uses the \collier
\cite{Denner:2014gla,Denner:2016kdg} library to calculate the one-loop
scalar
\cite{'tHooft:1978xw,Beenakker:1988jr,Dittmaier:2003bc,Denner:2010tr}
and tensor integrals
\cite{Passarino:1978jh,Denner:2002ii,Denner:2005nn} numerically.  The
complex-mass scheme \cite{Denner:1999gp,Denner:2005fg} is used
throughout.

\paragraph{Real radiation:}

The real QCD corrections are obtained by attaching a gluon in all
possible ways to the Born processes listed in
Eq.~\eqref{eq:bornprocesses}.  Consequently, one has to consider
four types of processes with all possible quark-flavour combinations:
\begin{align}
\label{eq:realqg}
 \Pg\Pg &\to \mu^-\bar{\nu}_\mu\Pb\bar{\Pb} q_l \bar q_m \Pg, \nonumber \\
 q_i \bar q_i/ \bar q_i q_i &\to \mu^-\bar{\nu}_\mu\Pb\bar{\Pb} q_l \bar q_m \Pg, \nonumber \\
 q_i \Pg/ \Pg q_i &\to \mu^-\bar{\nu}_\mu\Pb\bar{\Pb} q_l \bar q_m q_i, \nonumber \\
 \bar q_i \Pg/ \Pg \bar q_i &\to \mu^-\bar{\nu}_\mu\Pb\bar{\Pb} q_l \bar q_m \bar q_i,
\end{align}
with $q_i \in \{ \Pu, \Pd, \Pc, \Ps \}$ and $q_l q_m \in \{ \Pu \Pd,
\Pc \Ps \}$.  
Moreover, real photon radiation to the interferences of order $\order{g_s^2 g^4}$ and $\order{g_s^4 g^2}$ contributions have to be taken into account in order to ensure IR finiteness of the
corrections of order $\order{\alphas^3\alpha^{4}}$.
Thus, the following processes have to be included:
\begin{align}
 \Pg\Pg &\to \mu^-\bar{\nu}_\mu\Pb\bar{\Pb} q_l \bar q_m \gamma, \nonumber \\
 q_i \bar q_i/ \bar q_i q_i &\to \mu^-\bar{\nu}_\mu\Pb\bar{\Pb} q_l \bar q_m \gamma,
\end{align}
with $q_i \in \{ \Pu, \Pd, \Pc, \Ps \}$ and $q_l q_m \in \{ \Pu \Pd,
\Pc \Ps \}$.  

\changed{Note that the photon-induced real corrections of order
  $\order{\alphas^3\alpha^{4}}$ resulting from $q_i \gamma / \gamma
  q_i \to \mu^-\bar{\nu}_\mu\Pb\bar{\Pb} q_l \bar q_m q_i$ and $\Pg
  \gamma / \gamma q_i \to \mu^-\bar{\nu}_\mu\Pb\bar{\Pb} q_l \bar
  q_m\Pg$ have been neglected.  They are suppressed owing to the
  photon PDF and the fact that in the resulting
  $\order{\alphas^3\alpha^{4}}$ contributions at least one of the
  amplitudes does not feature doubly-resonant top contributions. They
  are expected to be smaller than the LO photon-induced contributions
  discussed at the end of \refse{sec:definition} and therefore
  negligible.}

To handle the IR singularities in the real contributions, the dipole
subtraction method \cite{Catani:1996vz,Dittmaier:1999mb} for both QCD
and QED has been used.  The colour-correlated matrix elements have
been obtained from the computer code \recola.  All singularities (both
of QCD and QED origin) from collinear initial-state splittings have
been absorbed in the PDFs using the $\overline{\text{MS}}$
factorisation scheme.

\paragraph{Validation:}

In order to ensure the validity of the calculation, several checks
have been performed.  The LO hadronic cross section has been compared
against the program {\sc\small
  MadGraph5\_aMC@NLO}~\cite{Alwall:2014hca}.  In order to verify the
IR and ultra-violet (UV) finiteness, the corresponding regulators have
been varied.  For each representative partonic channel, the cross
sections as well as representative distributions turn out to be
independent of such variations.  To check the implementation of the
subtraction mechanism, the $\alpha$ parameter\footnote{The present
  computation has been done using the value $\alpha= 10^{-2}$.}
\cite{Nagy:1998bb} has been changed from $10^{-2}$ to 1.  This
parameter restricts the dipole subtraction terms to the vicinity of
the singular regions and should drop out of the final results after
the inclusion of the corresponding integrated dipoles.  This has been
checked at both the level of the fiducial cross section and
differential distributions.  A Ward identity for the gg channel has
been verified by calculating for 4000 phase-space points the quantity
$\Re \left[\mathcal{M}^*_0(\epsilon_\Pg)\mathcal{M}_1(\epsilon_\Pg\to
  p_\Pg/p^0_\Pg)\right]/\Re\left[\mathcal{M}^*_0(\epsilon_\Pg)\mathcal{M}_1(\epsilon_\Pg)\right]$
where one of the initial gluons' polarisation vector $\epsilon_\Pg$
has been replaced by its momentum $p_\Pg$ normalised to its energy
$p^0_\Pg$.  Computing the cumulative fraction of events above certain
thresholds gives results comparably good as the ones of
\citeres{Denner:2015yca,Denner:2016jyo,Denner:2016wet}.  The one-loop
matrix elements obtained from \recola have been compared against the
ones of {\sc Recola}2 \cite{Denner:2017wsf} where a background-field
method \cite{Denner:2017vms} formulation of the Standard Model has
been implemented.  For 4000 phase-space points, it always gave at
least 4 digits agreement.

\section{Numerical analysis}
\label{sec:results}

\subsection{Input parameters}
\label{subsec:input}

The results presented here are for the LHC running at a centre-of-mass
energy of $\sqrt{s}=13\TeV$.  To interface the PDFs, the program
LHAPDF 6.1.5~\cite{Andersen:2014efa,Buckley:2014ana} has been
utilised.  We have used the
$\mathrm{NNPDF30}\_\{\mathrm{lo}/\mathrm{nlo}\}\_\mathrm{as}\_0118$
PDF sets \cite{Ball:2014uwa} at LO and NLO, respectively.  The central
value of the factorisation and renormalisation scale has been chosen
to be
\begin{equation}
\label{eq:scale}
\mu_0 = \Etbar / 2 = \frac12 \sqrt{\sqrt{\Mt^2+\ptsub{\Pt}^2} \sqrt{\Mt^2+\ptsub{\bar \Pt}^2}} ,
\end{equation}
with $\ptsub{\bar \Pt}/\ptsub{\Pt}$ standing for the transverse momentum of the top/antitop quark.
This choice is motivated by previous computations of off-shell
top--antitop production in the fully leptonic channel
\cite{Denner:2010jp,Denner:2012yc}.  It leads to small
scale dependencies and moderate NLO corrections (see
\reffi{plot:scale} and related discussion below),  
as further discussed in \refses{sec:crossection} and
\ref{sec:diffdist}.

The electromagnetic coupling $\alpha$ has been fixed by the Fermi constant in the $G_\mu$ scheme \cite{Denner:2000bj} as
\begin{equation}
  \alpha = \frac{\sqrt{2}}{\pi} G_\mu \MW^2 \left( 1 - \frac{\MW^2}{\MZ^2} \right), 
  \qquad \text{with}  \qquad   \GF    = 1.16637\times 10^{-5}\GeV^2.
\end{equation}
The numerical values of the masses and widths read \cite{Patrignani:2016xqp}:
\begin{alignat}{2} 
                 \Mt   &=  173.34\GeV,       & \quad \quad \quad  M_{\rm H} &=  125.0\GeV,  \nonumber \\
                \MZOS &=  91.1876\GeV,      & \GZOS &= 2.4952\GeV, \nonumber \\
                \MWOS &=  80.385\GeV,       & \GWOS &= 2.085\GeV ,
\end{alignat}
with the Higgs-boson mass taken following the recommendations of
\citere{deFlorian:2016spz}.  The bottom quark is considered
massless.  The pole masses and widths entering the calculation are
determined from the measured on-shell (OS) values \cite{Bardin:1988xt}
for the W and Z~bosons according to
\begin{equation}
M_V = \frac{\MVOS}{\sqrt{1+(\GVOS/\MVOS)^2}},\qquad  
\Gamma_V = \frac{\GVOS}{\sqrt{1+(\GVOS/\MVOS)^2}}.
\end{equation}
The mass and width of the top quark are taken from
\citere{Basso:2015gca}, where $\Gt^\text{LO} = 1.449582\GeV$ at LO
and $\Gt^\text{NLO} = 1.35029\GeV$ at NLO QCD, respectively.

\subsection{Event selection}
\label{subsec:cuts}

The event selection is inspired by the searches performed at the LHC
by the ATLAS and CMS collaborations in the lepton+jets channel
\cite{Aaboud:2017fha,Khachatryan:2016mnb}.  The jets (light as well as
bottom jets) are clustered with the anti-$k_\text{T}$ algorithm
\cite{Cacciari:2008gp} using a jet radius of $R=0.4$.  A bottom jet
clustered with a light jet gives rise to a bottom jet.  Note that for
photon recombination with charged particles, the clustering radius is
taken to be $R=0.1$.  The event selection for the final state reads:
\begin{alignat}{5} 
\label{eq:cuts}
                 \text{light/bottom jets:}                     && \qquad \ptsub{\Pj/\Pb}         &>  25\GeV,  & \qquad |y_{\Pj/\Pb}|   &< 2.5, & \nonumber \\
                \text{charged lepton:}              && \ptsub{\Pl}         &>  25\GeV,  & |y_{\Pl}| &< 2.5, &
\end{alignat}
with $y$ standing for the rapidity.
The final state is thus characterised by two light jets, two bottom jets, a charged lepton, and missing energy.
This implies that effectively the jet radius is acting as a cut,
\begin{equation}
\label{eq:Rcuts}
 \Delta R_{\Pj\Pj}, \Delta R_{\Pj\Pb}, \Delta R_{\Pb\Pb} > 0.4,
\end{equation}
where the distance between two particles $i$ and $j$ is defined as
\begin{equation}
\label{eq:distance}
        \Delta R_{ij} = \sqrt{(\Delta \phi_{ij})^2+(\Delta y_{ij})^2},
\end{equation}
with the azimuthal angle difference $\Delta
\phi_{ij}=\min(|\phi_i-\phi_j|,2\pi-|\phi_i-\phi_j|)$.

This set of cuts aims at measuring top-pair production in the
\emph{resolved} topology as opposed to the \emph{boosted}
topology.\footnote{Such a distinction is for example made by the ATLAS
  collaboration in \citere{Aaboud:2017fha}.}  In the resolved event
selection, the decay products of the hadronically decaying top quark
are required to be separated.  The boosted selection, on the other
hand, is used for measurements of top quarks with large momenta in
association with large-$R$ jets.  In order to reduce the
non-$\Pt\bar{\Pt}$ background, we have required that at least one
jet--jet invariant mass fulfils the criterion
\begin{equation}
\label{eq:mjjcuts}
 60 \GeV < m_{\Pj\Pj} < 100 \GeV .
\end{equation}
Hence the two jets are most probably originating from the decay of a
W~boson and thus of a top quark.  This ensures that the bulk of the
cross section is originating mainly from two resonant top quarks and not
from background contributions.  In particular, it removes real
radiation events where the two jets originating from the W-boson decay
are recombined into a single jet, while the extra real radiation
gives rise to the presence of a second separated jet.  Such events typically
have boosted kinematics and can make the real contribution potentially
very large.  This effect of quarks being recombined at high
transverse momentum has already been foreseen in \citere{Baur:2006sn}.

\subsection{Cross sections}
\label{sec:crossection}

In \refta{table:crossection}, the fiducial cross sections at both LO
and NLO are presented.  At LO, the $\Pg \Pg$ channel amounts to
$90.0\%$ of the fiducial cross section while the ${ q \bar{q}}$ ones
account for $10.0\%$.  
\changed{ The contributions with bottom quarks
  in the initial state turn out to be completely negligible and are
  around a per mille ($0.13\%$ of the LO cross section) which is the
  Monte Carlo error of our NLO calculation.}  Therefore, these
contributions have been omitted in the predictions presented here.  At
NLO, the $\Pg \Pg$ channel represents $83.7\%$ of the fiducial cross
section while the ${ q \bar{q}}$ one is amounting to $6.0\%$.  At NLO,
the real corrections with gluon--quark initial states in
Eq.~\eqref{eq:realqg} are accounting for $10.3\%$ of the cross
section.
\begin{table}
\begin{center} 
\begin{tabular}{ c  c  c   c }
 Ch. & $\sigma_{\rm LO}$ [pb] & $\sigma_{\rm NLO}$ [pb] & $K$-factor\\
  \hline\hline
$\Pg \Pg$             & $12.0257(5)$  & $13.02(7)$  & $1.08$ \\
${ q \bar{q}}$        & $1.3308(3)$  & $0.942(7)$   & $0.71$ \\
${\rm g} q(/\bar{q})$ &   & $1.604(5)$  &        \\
  \hline
$\Pp\Pp$              & $13.3565(6)$ & $15.56(7)$  & $1.16$ \\
  \hline
\end{tabular}
\end{center}
\caption{
Fiducial cross sections at LO and NLO for the process $\Pp\Pp\to
\mu^-\bar{\nu}_\mu\Pb\bar{\Pb} \Pj \Pj$ with its corresponding
sub-channels. 
                The possible flavours of the quark are $ q=\Pu,\Pd,\Pc,\Ps$.
                The quark-gluon channels denoted by ${\rm g} q(/\bar{q})$ consist in the real corrections with gluon--quark initial states in Eq.~\eqref{eq:realqg} and appear only at NLO.
                The proton--proton cross section is presented in the last line of the table dubbed $\Pp\Pp$.
                The $K$-factors are defined as $K = \sigma_{\rm NLO} / \sigma_{\rm LO}$.
                The integration errors of the last digits are given in parentheses.
                The predictions are expressed in pb and are for the LHC running at a centre-of-mass energy of $\sqrt{s}=13\TeV$.}
\label{table:crossection}
\end{table}

It is worth noting that the $K$-factor of the gluonic channel is
larger than the one for quark--antiquark annihilation, amounting to
$1.08$ and $0.71$, respectively.  Different $K$-factors for different
partonic channels have already been observed for similar processes
\cite{Bredenstein:2008zb,Bredenstein:2009aj,Bredenstein:2010rs,Denner:2010jp,Denner:2012yc,Denner:2015yca}.
Owing to the addition of the ${\rm g} q(/\bar{q})$ channels of
Eq.~\eqref{eq:realqg} at NLO, the $K$-factor of the fiducial cross
section is $1.16$.

The effect of the variation of the factorisation and renormalisation
scales on the total prediction has been studied.  To this end, the
central value $\mu= \Etbar/2$ has been re-scaled by factors $\xi_{\rm
  fac}$ and $\xi_{\rm ren}$ for 
\begin{align}
\label{eq:combination}
 \left(\xi_{\rm fac},\xi_{\rm ren}\right) \in \left\{\left(1/2,1/2\right),\, \left(1/2,1\right),\, \left(1,1/2\right),\, \left(1,1\right),\, \left(1,2\right),\, \left(2,1\right),\, \left(2,2\right)\right\},
\end{align}
where $\left(\xi_{\rm fac},\xi_{\rm ren}\right) = \left(1,1\right)$
represents the central scale.  In addition to the  cross sections
for the central scale reported in \refta{table:crossection}, the
lowest and highest cross sections for the combinations of
Eq.~\eqref{eq:combination} have been extracted.  The fiducial cross
sections with scale variation read
\begin{equation} 
\sigma_{\rm LO} =  13.3565(6)_{-22.09\%}^{+30.68\%}\pb 
\end{equation}
and
\begin{equation} 
\label{eq:NLO}
\sigma_{\rm NLO} =  15.56(7)_{-4.6(5)\%}^{+0.9(6)\%}\pb ,
\end{equation}
where the values in per cent represent the extrema of the cross
sections calculated for the scales \eqref{eq:combination}.  As
expected, there is a significant reduction of the scale uncertainty
(more than a factor four) when going from LO to NLO accuracy.
\changed{The asymmetric scale uncertainty at NLO is due to our choice
  of the central scale near the maximum of the cross section. The
  positive uncertainty in Eq.~\eqref{eq:NLO} results in fact from
  off-diagonal scale variations.}

\begin{figure}
\begin{center}
         \includegraphics[width=0.7\linewidth]{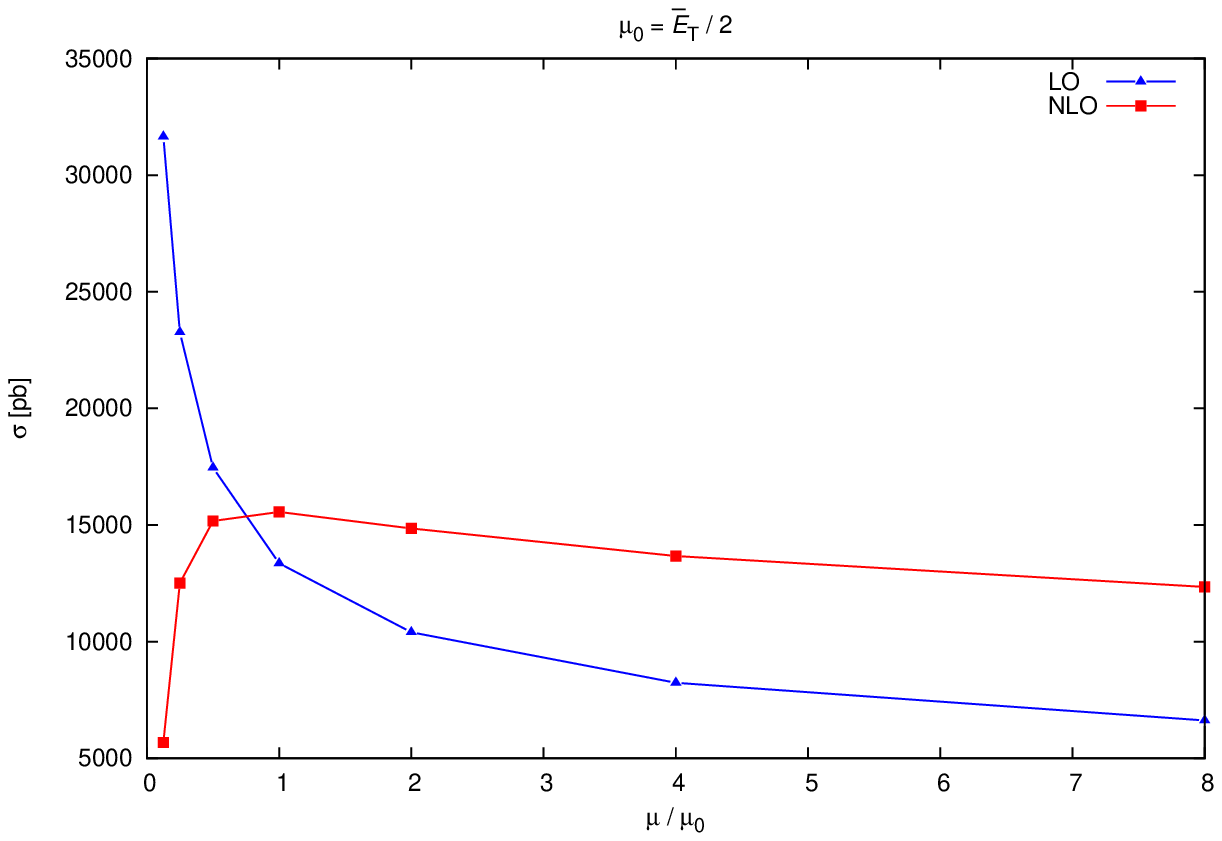}
\end{center}
        \caption{%
                Scale dependence of the LO and NLO cross sections at a centre-of-mass energy $\sqrt{s}=13\TeV$ at the LHC for $\Pp\Pp\to \mu^-\bar{\nu}_\mu\Pb\bar{\Pb} \Pj \Pj$.
                The renormalisation and factorisation scales are varied together around the central scale $\mu_0 = \Etbar / 2$ defined in Eq.~\eqref{eq:scale}.
        }
\label{plot:scale}
\end{figure}

In \reffi{plot:scale} the fiducial cross section at both LO and NLO
accuracy is given as a function of the ratio of scales $\mu / \mu_0$
in the range $[1/8,\;8]$.  Both the factorisation and renormalisation
scales are set equal to the scale $\mu$, while the scale $\mu_0 =
\Etbar / 2$ is defined in Eq.~\eqref{eq:scale}.  The LO cross section
shows the usual unbounded exponential behaviour, while the NLO
prediction displays a much smaller scale dependence.  Near $\mu_0 =
\Etbar / 2$ the NLO cross section is flat, and the scale variation is
minimal.  Hence, the choice $\mu=\mu_0$ ensures a maximal reduction of
the scale uncertainty when going from LO to NLO for the fiducial cross
section. Note however, that the resulting small positive scale
variation tends to underestimate the uncertainty in this case
\changed{and the choice $\mu_0 =\Etbar$ would provide a more
  conservative NLO scale uncertainty.  While for $\mu_0 = \Etbar/ 2$
  the difference between LO and NLO predictions is reasonably small,
  guaranteeing that the NLO cross section is within LO scale
  uncertainty, this is not the case for the choice $\mu_0 = \Etbar$
  which leads to much larger NLO corrections. In any case, the NLO
  cross section has a good perturbative behaviour and provides a much
  more reliable prediction.}

\changed{ Finally, we recall that this computation involves only the
  partonic channels that feature two resonant top quarks and thus two
  resonant W~bosons [see
  Eq.~\eqref{eq:bornprocesses}].  All other partonic channels receive
  only contributions with one resonant top quark and one resonant W~boson.
  Imposing a cut on the invariant mass of the two jets
  around the W-boson mass additionally suppresses these contributions 
  both at the level of the cross section and differential
  distributions.\footnote{\changed{In the next section, comments are made for
    differential distributions where the effect of non-doubly-top-resonant
    partonic channels is larger than $1\%$.}}  It thus renders these
  partonic channels phenomenologically negligible.}

\subsection{Differential distributions}
\label{sec:diffdist}

Turning to the differential distributions, for each of them, the
LO and NLO predictions are shown in the upper plot while the ratio of
the two predictions is presented in the lower panel.  The band is
obtained by variation of the factorisation and renormalisation scales
independently within the set of Eq.~\eqref{eq:combination}.  In the
NLO/LO ratio, the predictions are normalised to the LO ones for the
central scale.

\paragraph{Transverse-momentum distributions.}

\begin{figure}
        \setlength{\parskip}{-10pt}

        \begin{subfigure}{0.49\textwidth}
                \subcaption{}
                \includegraphics[width=\textwidth]{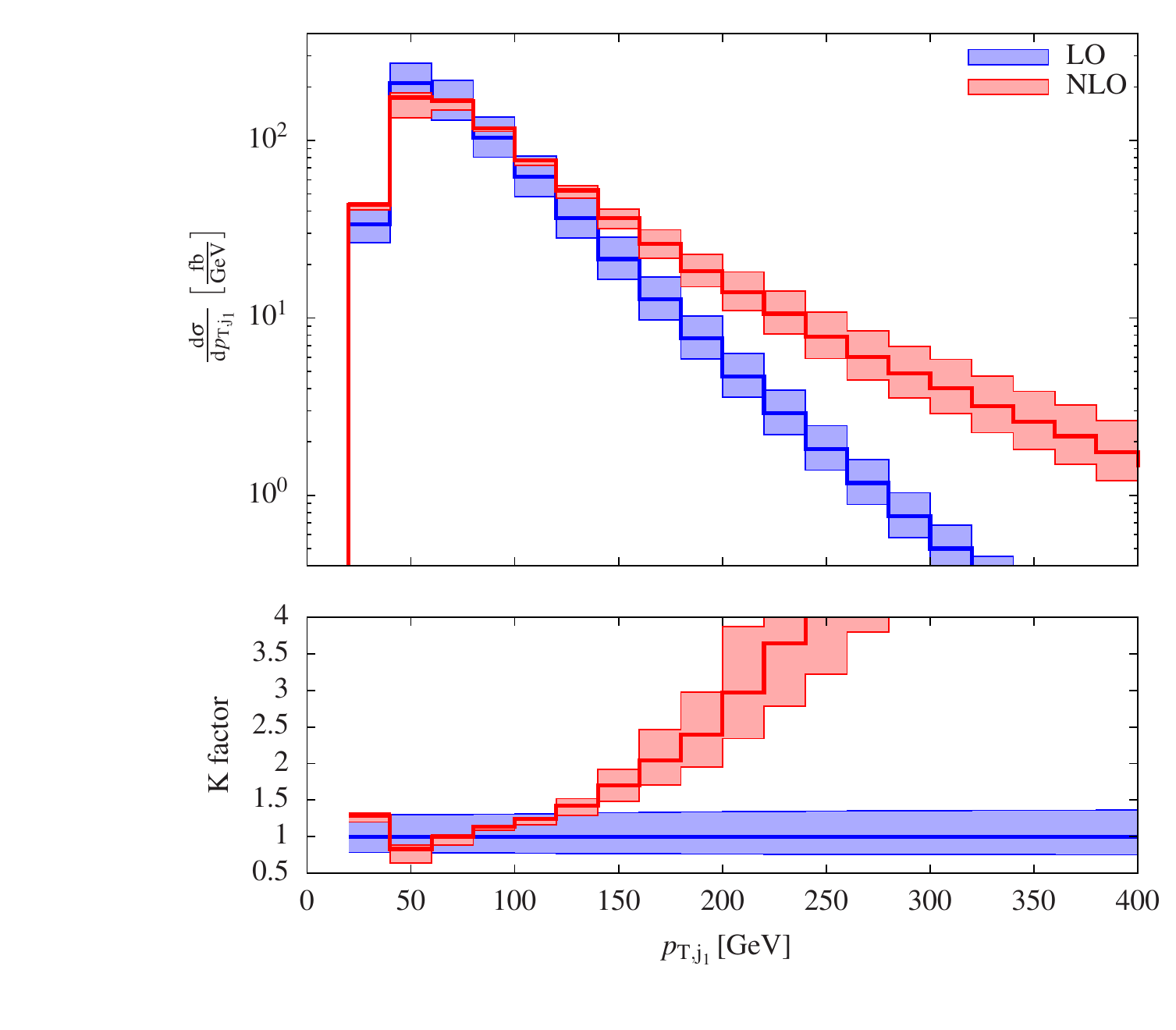}
                \label{plot:pt_j1} 
        \end{subfigure}
        \hfill
        \begin{subfigure}{0.49\textwidth}
                \subcaption{}
                \includegraphics[width=\textwidth]{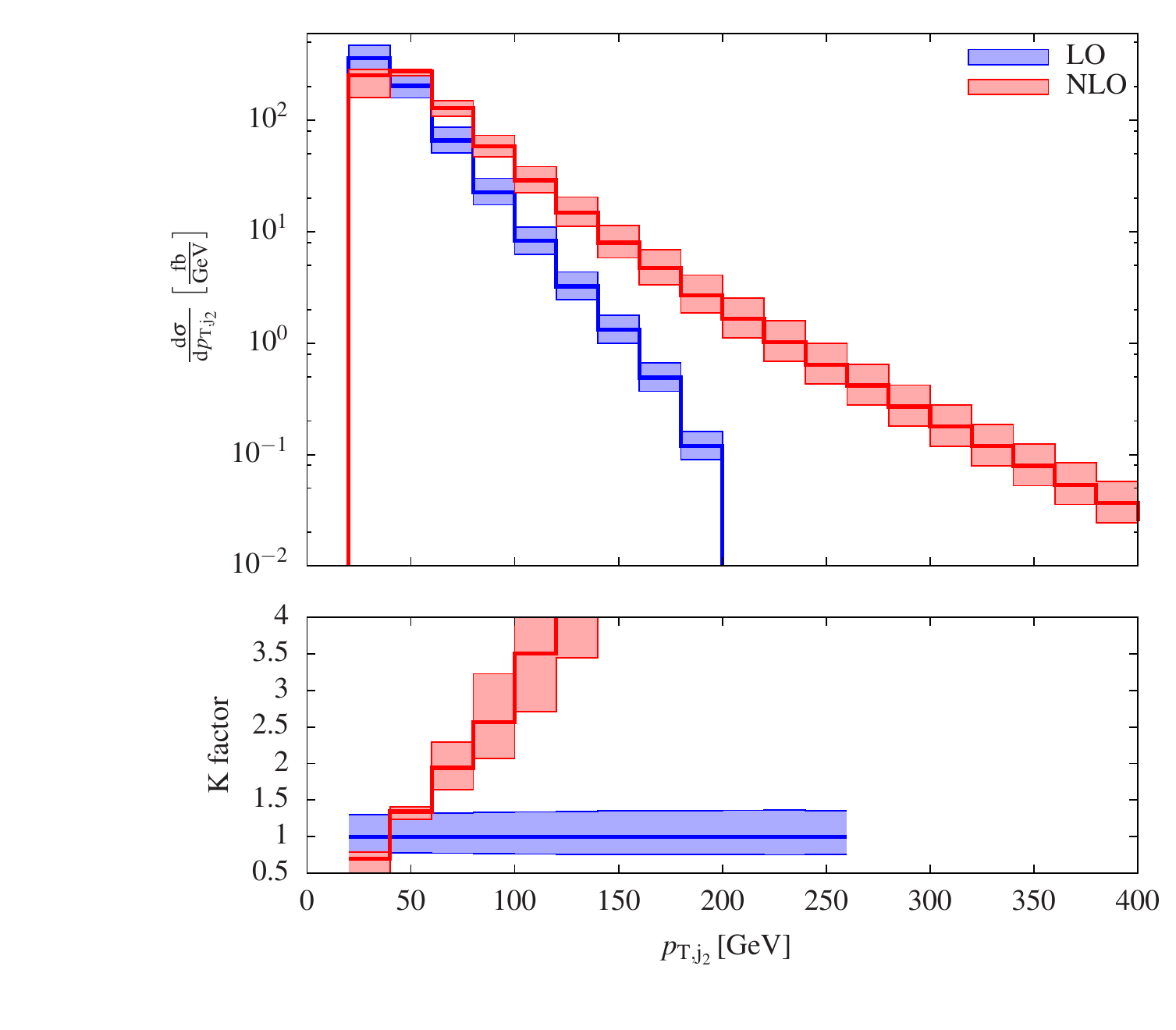}
                \label{plot:pt_j2}
        \end{subfigure}
        
        \begin{subfigure}{0.49\textwidth}
                \subcaption{}
                \includegraphics[width=\textwidth]{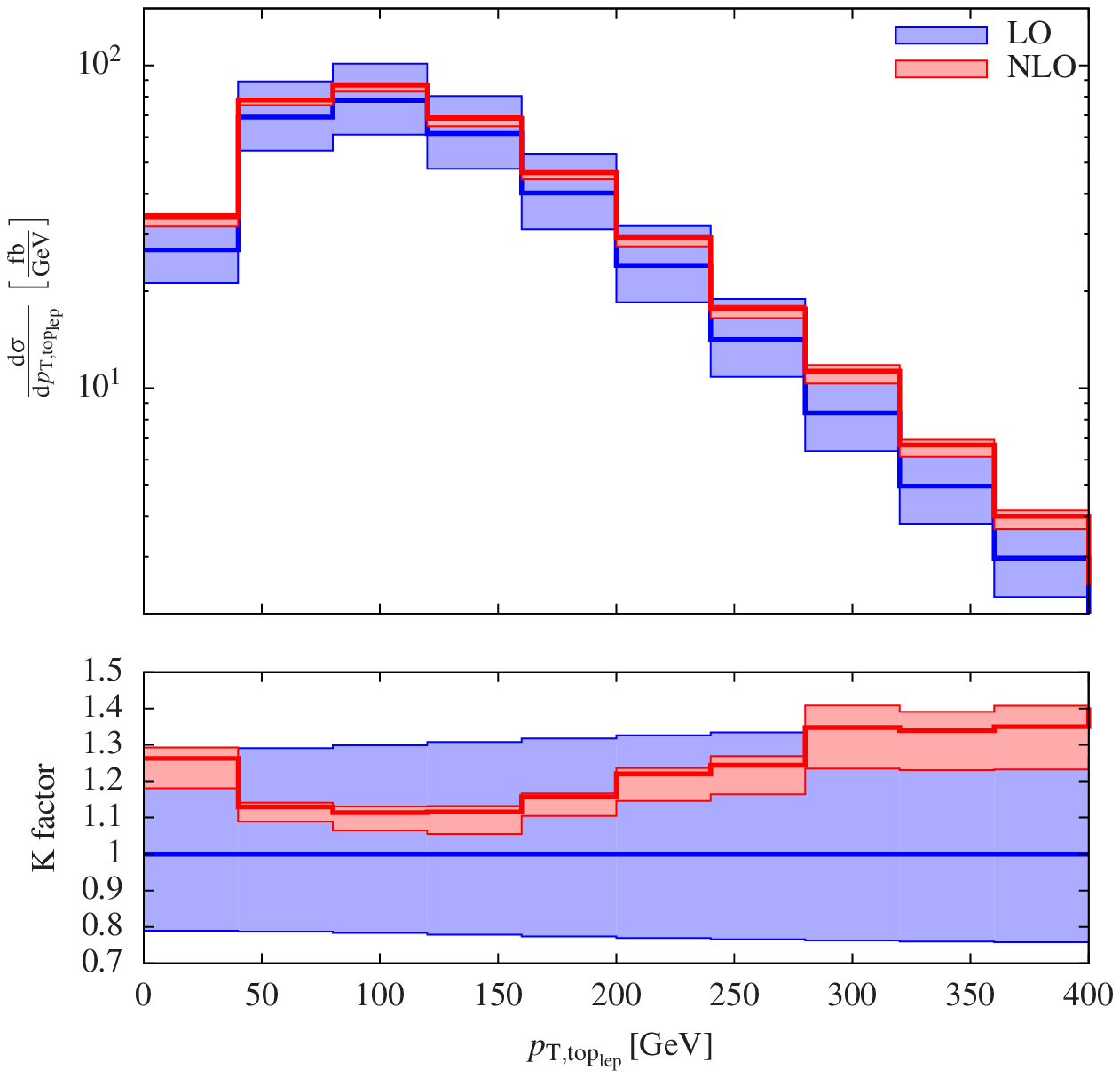}
                \label{plot:pt_top_lep}
        \end{subfigure}
        \hfill
        \begin{subfigure}{0.49\textwidth}
                \subcaption{}
                \includegraphics[width=\textwidth]{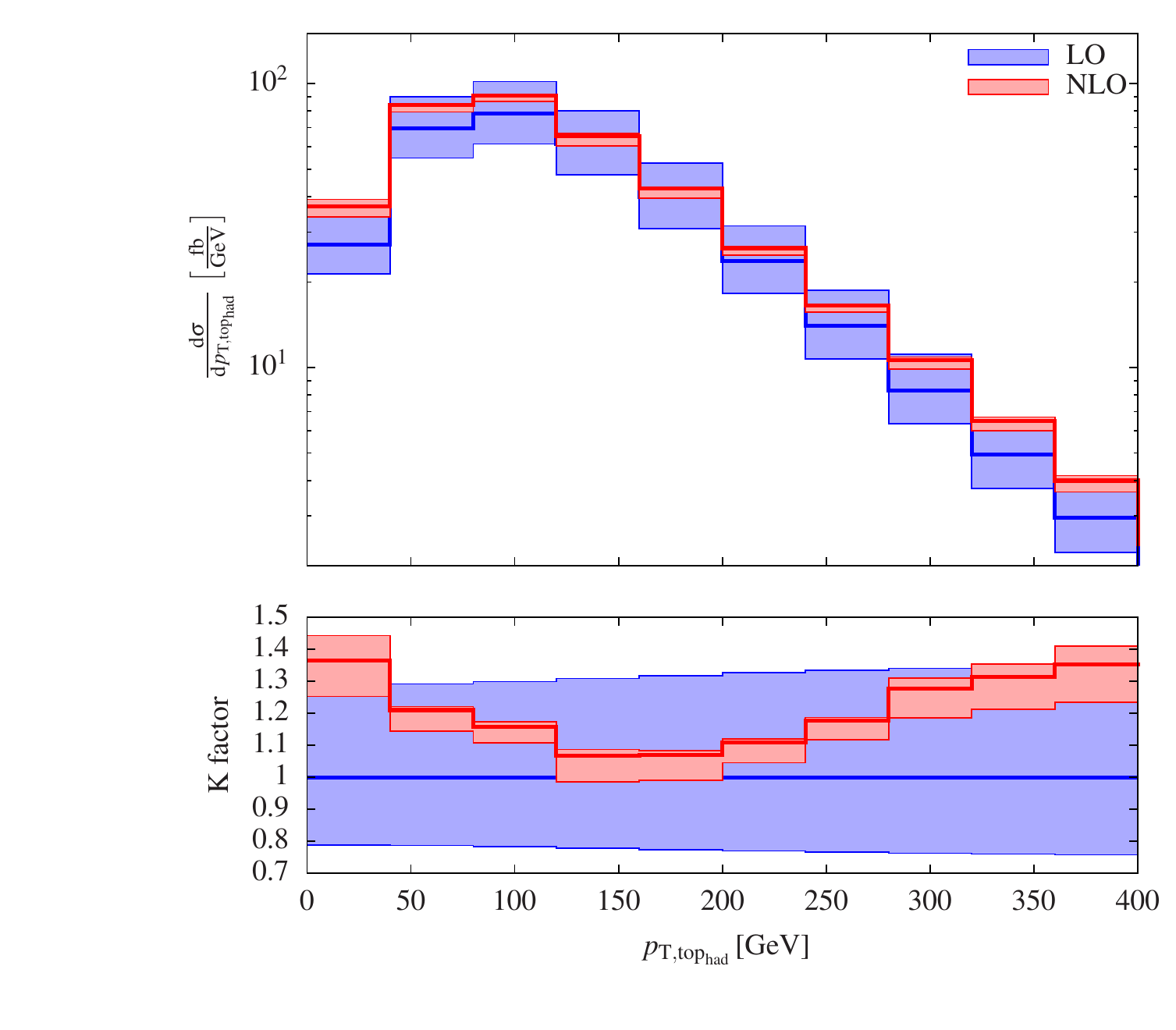}
                \label{plot:pt_top_had}
        \end{subfigure}
        
        \begin{subfigure}{0.49\textwidth}
                \subcaption{}
                \includegraphics[width=\textwidth]{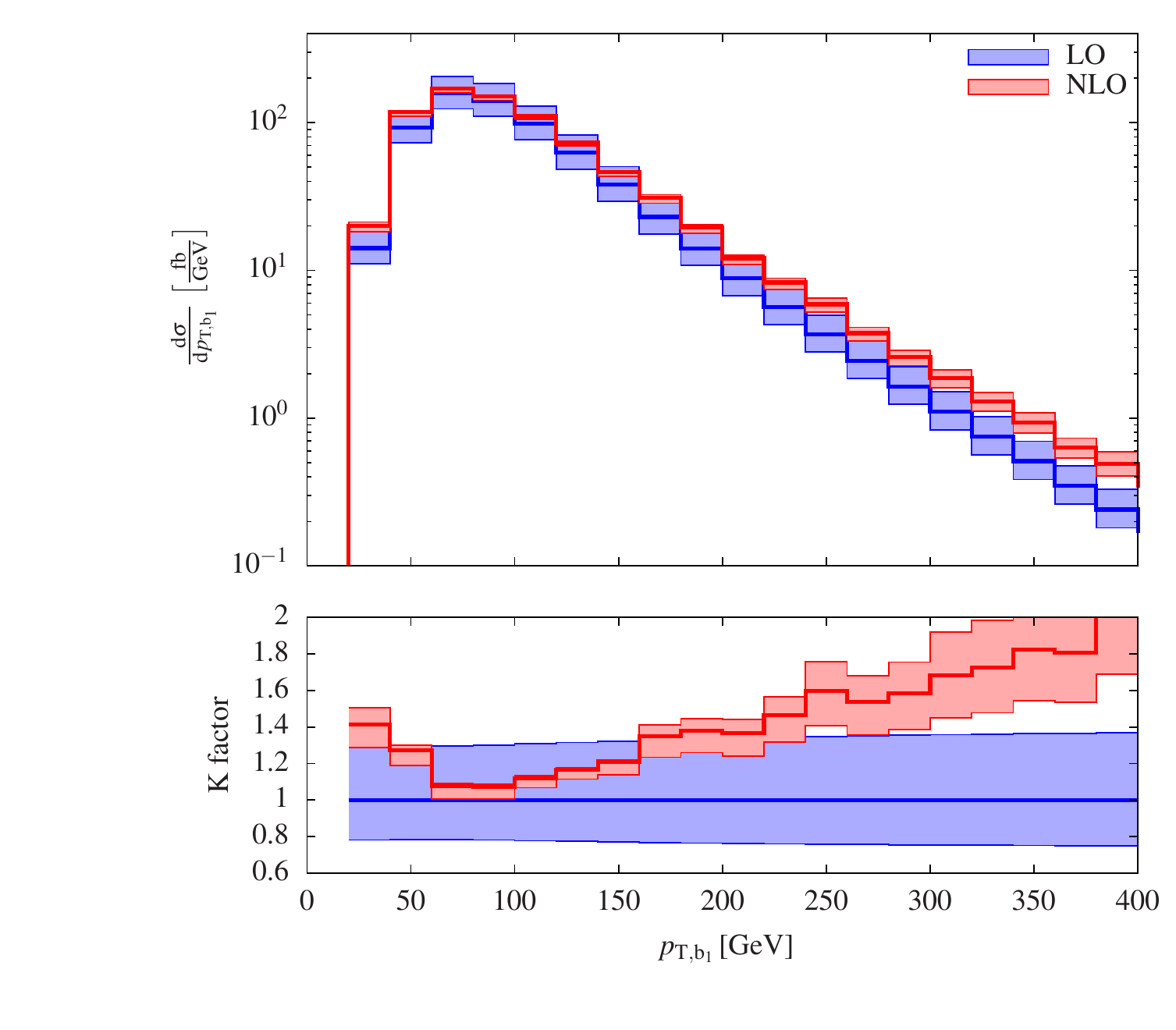}
                \label{plot:pt_b1}
        \end{subfigure}
        \hfill
         \begin{subfigure}{0.49\textwidth}
                \subcaption{}
                \includegraphics[width=\textwidth]{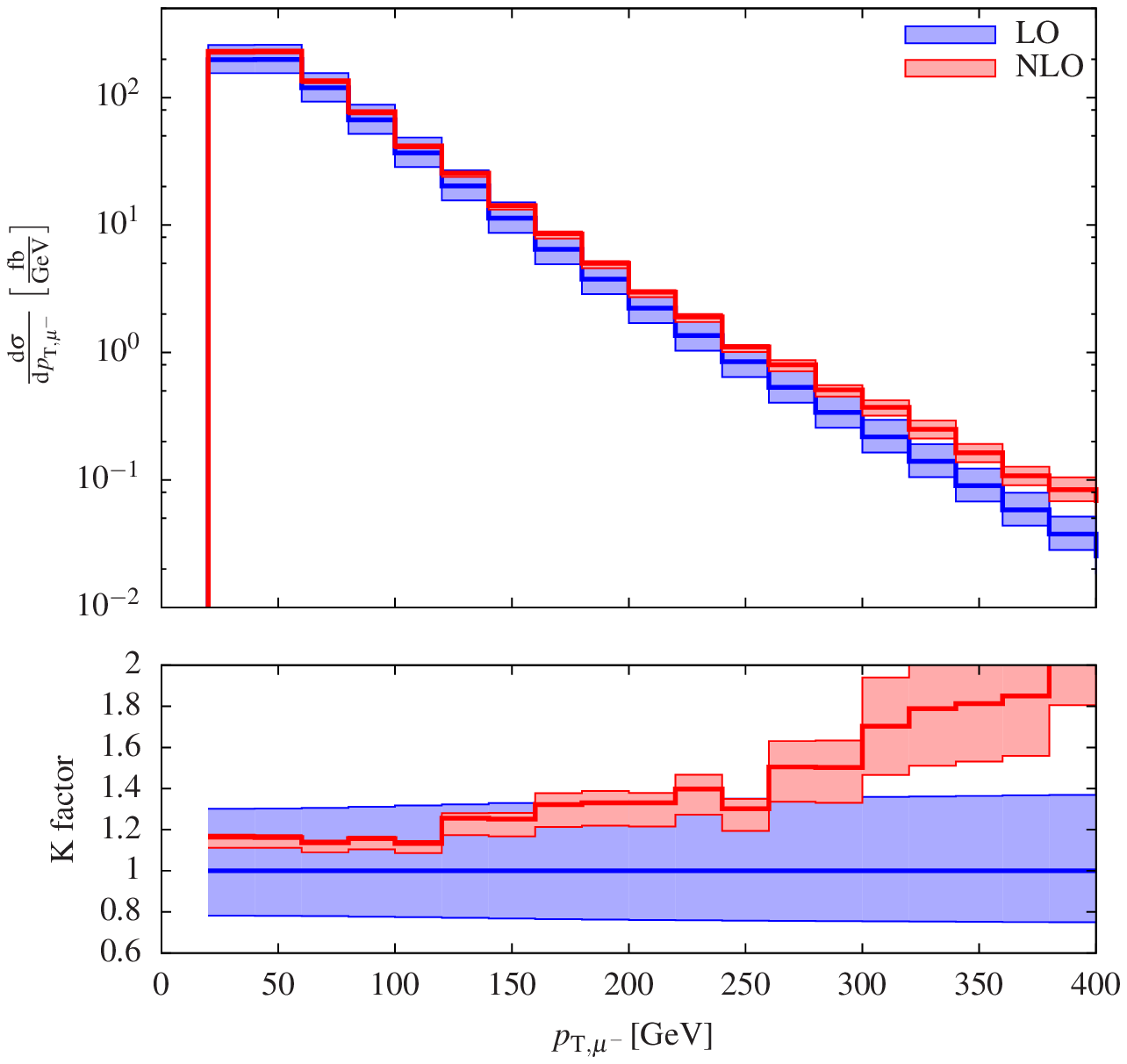}
                \label{plot:pt_mu}
        \end{subfigure}
 
        \vspace*{-5ex}
        \caption{\label{fig:pt_distributions}%
                Differential distributions at a centre-of-mass energy $\sqrt{s}=13\TeV$ at the LHC for $\Pp\Pp\to \mu^-\bar{\nu}_\mu\Pb\bar{\Pb} \Pj \Pj$: 
                \subref{plot:pt_j1}~transverse momentum of hardest jet~(top left), %
                \subref{plot:pt_j2}~transverse momentum of the second hardest jet~(top right),
                \subref{plot:pt_top_lep}~transverse momentum of the leptonic top quark~(middle left),
                \subref{plot:pt_top_had}~transverse momentum of the reconstructed hadronic top quark~(middle right),
                \subref{plot:pt_b1}~transverse momentum of the hardest
                bottom jet~(bottom left), and
                \subref{plot:pt_mu}~transverse momentum of the muon~(bottom right).
                The upper panels show the LO prediction as well as the NLO one.
                The lower panels display the ratio of the NLO and the
                LO predictions. The bands correspond to factor-2 scale
                variations as defined in Eq.~\eqref{eq:combination}.}
\end{figure}
In \reffi{fig:pt_distributions}, several transverse-momentum
distributions are shown.  The first two, in \reffis{plot:pt_j1} and
\ref{plot:pt_j2}, are the ones for the hardest and second hardest
jet,\footnote{The hardest jet is the one with the highest transverse
  momentum, etc.} respectively.  In \reffi{plot:pt_j1}, one observes a
strong increase of the NLO corrections towards high transverse
momentum.  Below $150\GeV$, the corrections stay below $100\%$ while
above they become more than one order of magnitude larger.  At
$400\GeV$, the NLO prediction is $31$ times larger than the LO
prediction.  This is a purely kinematical effect in combination with
the event selection which explains why the scale-variation band is not
a reliable estimate here.  Since the LO contribution is strongly
suppressed for high transverse jet momenta, the NLO scale-variation
band increases towards high transverse momentum.  There, the two jets
originating from the W~boson are collinear.  Consequently at LO such
configurations are forbidden due to the jet distance cut
\eqref{eq:Rcuts}, while at NLO such events are allowed owing to the
extra jet from real emission.  Hence these large corrections are due
to a suppression of the LO configuration in this phase-space region.
This effect is particularly explicit for the transverse momentum of
the second hardest jet.  At $200\GeV$, the LO predictions decrease
sharply to become even zero around $250\GeV$ because of the
invariant-mass cut and the $\Delta R$ cut.  Indeed, assuming small
angles between the two jets leads because of
${\ptsub{\Pj_2}}<{\ptsub{\Pj_1}}$ to $\ptsub{\Pj_2,\rm max}^2 \sim
{m_{\Pj \Pj,\rm max}^2} / \Delta R_{\Pj\Pj,\rm min}^2 = \left( 100
\right)^2 / \left( 0.4 \right)^2 = \left( 250 \GeV \right)^2$.  A
similar behaviour has already been observed in \citere{Denner:2014wka}
when considering doubly top-resonant contributions.  Above this
threshold, the NLO contributions consist exclusively of real radiation
where the two jets originating from the W-boson decay are recombined
in a single jet.  Such events are accepted if the invariant mass of
this jet and the real radiation jet is still fulfilling the
requirement of Eq.~\eqref{eq:mjjcuts}. Without this requirement, the
contribution of such real radiation events would be much larger.

The transverse momenta of the leptonic and hadronic top quarks are
displayed in \reffis{plot:pt_top_lep} and \ref{plot:pt_top_had},
respectively.  The definition of the leptonic top quark is based on
Monte Carlo truth and is thus the total momentum of the anti-bottom
quark (possibly recombined with a light jet), the charged lepton, and
the neutrino.  At low transverse momentum, the corrections are at the
level of $26\%$.  They are below $10\%$ at $150\GeV$ and increase
steadily to reach $35\%$ at $400\GeV$.  On the other hand, the
definition of the reconstructed hadronic top quark reads: from the two
or three light jets, retain the two momenta whose combination has the
invariant mass closest to the W-boson mass.  Out of the two bottom
quarks, retain the one whose momentum when added to the ones of the
two pre-selected jets results in a 3-jet invariant mass closest to the
top-quark mass.  The total momentum of these two light jets and bottom
jet defines the reconstructed hadronic top momentum.  The behaviour of
the corrections is comparable to the one for the leptonic top quark.
The corrections are around $37\%$ at zero transverse momentum, reach a
minimum at $150\GeV$, and are maximal at high transverse momentum
where the effect described above is taking place. At $400\GeV$, the
NLO corrections amount to $35\%$.  
\changed{This behaviour has already
  been observed for top-pair production with purely leptonic decays
  \cite{Denner:2012yc}. The increased corrections for small transverse
  momenta result from a redistribution of events where a gluon is
  emitted from one of the decay products of the top quark and carries
  away momentum. This effect is amplified in the case of hadronic top
  quark owing to the three jets in the final state that can radiate
  gluons while for the leptonic top quark, only one jet is present in
  the decay products.}
The transverse-momentum distributions for the top quarks are more
inclusive than the ones for the jets in the sense that they are built
from more momenta.  Therefore, they are less sensitive to the
kinematical effect described previously.  This explains why they are
slowly decreasing towards higher transverse momentum and do not
display extremely large corrections over the phase space studied.
Nonetheless the $K$-factor is not flat and becomes sizeable both for
low and high transverse momenta.

The distribution in the transverse momentum of the hardest bottom jet,
\reffi{plot:pt_b1}, exhibits a similar behaviour as the one in the
transverse momentum of the leading jet, \reffi{plot:pt_j1}, but much
less pronounced.  Again, the corrections increase in the tail of the
distribution where non-resonant contributions become relevant.  There,
the production of the $\mu^-\bar{\nu}_\mu\Pb\bar{\Pb} \Pj \Pj$ final
state proceeds increasingly through contributions that do not feature
two resonant top quarks \cite{Denner:2014wka,Denner:2016jyo}.  The
increase of the corrections for low transverse momentum is correlated
with the decrease of the LO predictions.  Finally, \reffi{plot:pt_mu}
displays the transverse momentum of the muon.  Apart from the small
transverse-momentum region, this observable shows a similar behaviour
as the transverse momentum of the hardest bottom jet.  The corrections
start below $20\%$ near the cut at $30\GeV$ to increase smoothly up to
a $K$-factor of $2.7$ at $400\GeV$.  This increase of the NLO
corrections towards high transverse momentum is also accompanied by an
increase of the size of the scale-variation band.

\changed{The contributions from non-doubly-top-resonant partonic
  channels are typically increasing towards high transverse momenta.
  For the distributions in the transverse momentum of the hardest jet
  and the hardest bottom jet,
  they exceed 1\% at $300\GeV$ and amount to $5\%$ at $500\GeV$.  For
  the distribution in the transverse momentum of the second hardest
  jet, they quickly increase up to $4\%$ around $200\GeV$, \ie above the kinematical
  threshold described above.  On the other hand, for the distributions
  in the transverse momentum of the top quarks (either leptonic or
  hadronic), these effects are below $1\%$ up to at least $500\GeV$ as
  these observables are more inclusive than those in the transverse
  momentum of single particles.}%

\paragraph{Invariant-mass distributions.}

\begin{figure}
        \setlength{\parskip}{-10pt}

        \begin{subfigure}{0.49\textwidth}
                \subcaption{}
                \includegraphics[width=\textwidth]{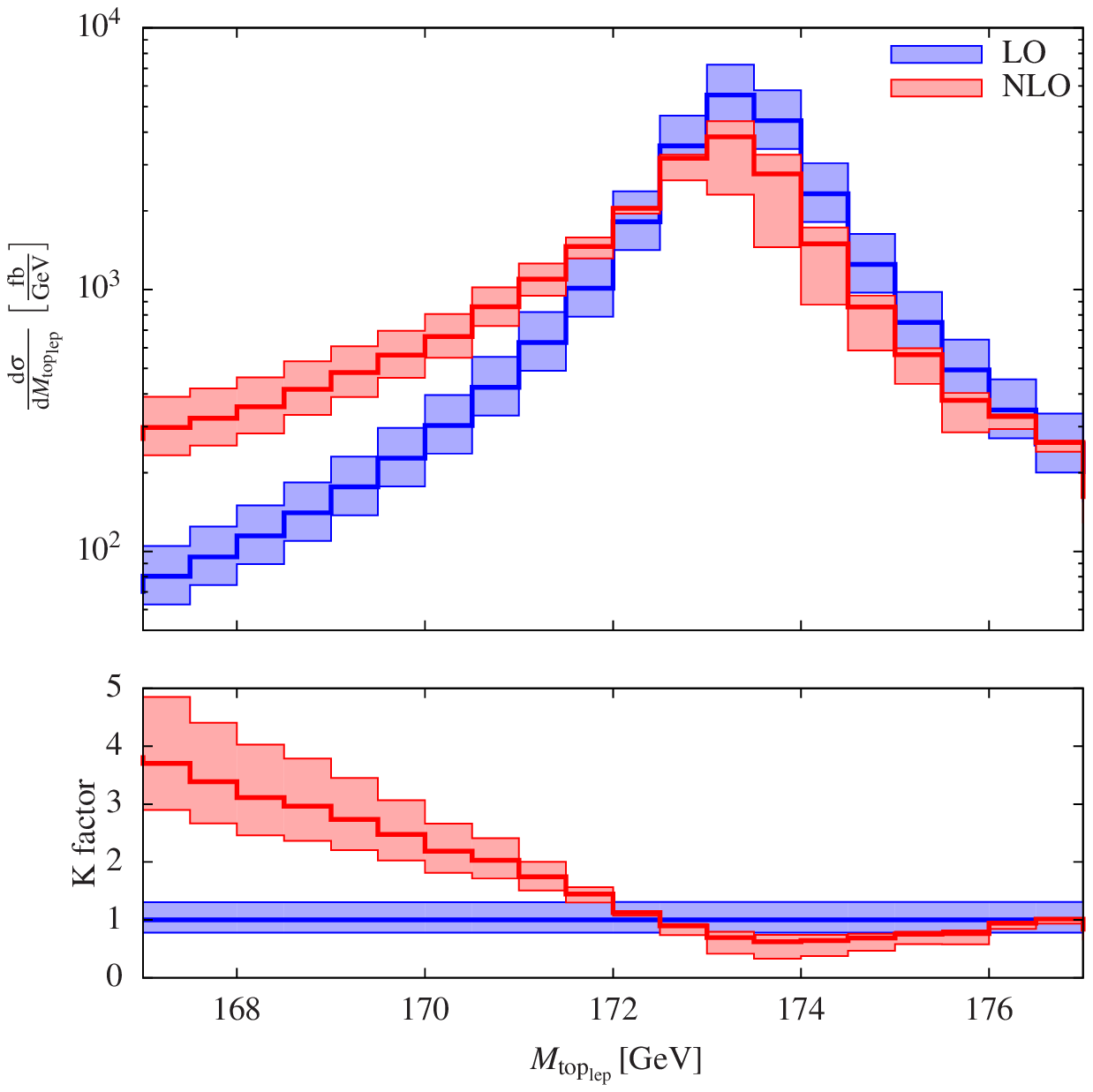}
                \label{plot:inv_antitop}
        \end{subfigure}
        \hfill
        \begin{subfigure}{0.49\textwidth}
                \subcaption{}
                \includegraphics[width=\textwidth]{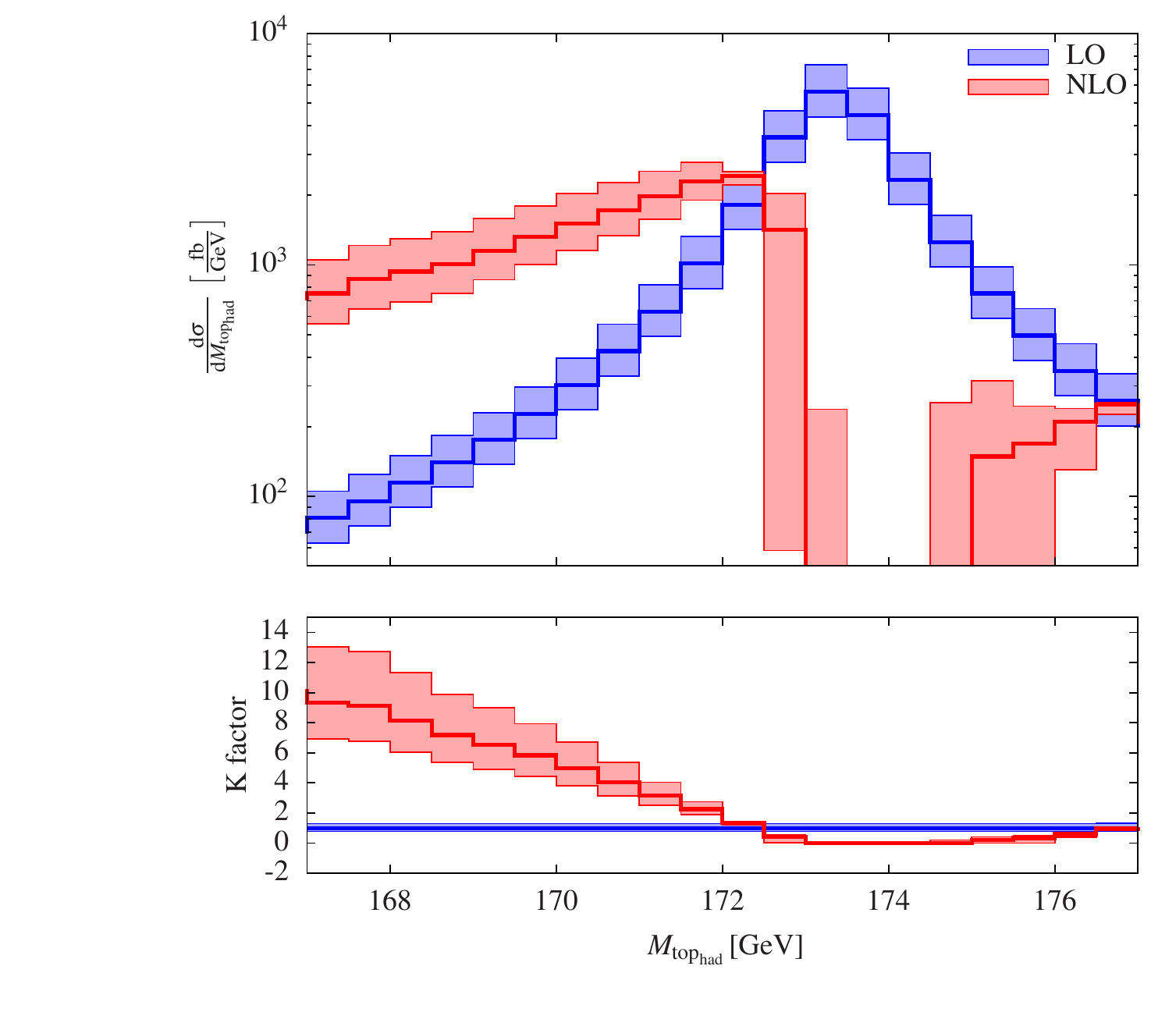}
                \label{plot:inv_tophad}
        \end{subfigure}

        \begin{subfigure}{0.49\textwidth}
                \subcaption{}
                \includegraphics[width=\textwidth]{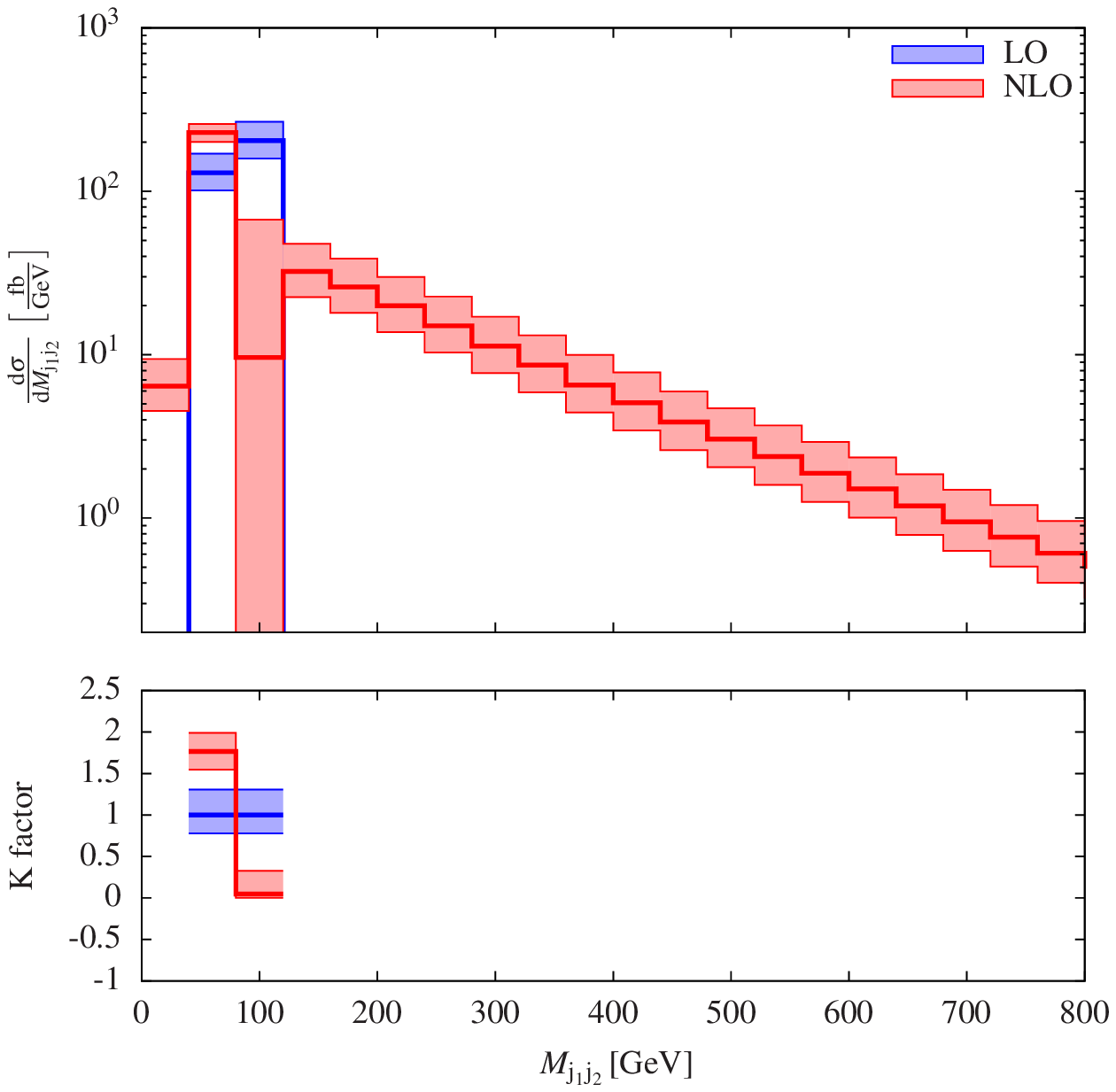}
                \label{plot:inv_jj}
        \end{subfigure}
        \hfill
        \begin{subfigure}{0.49\textwidth}
                \subcaption{}
                \includegraphics[width=\textwidth]{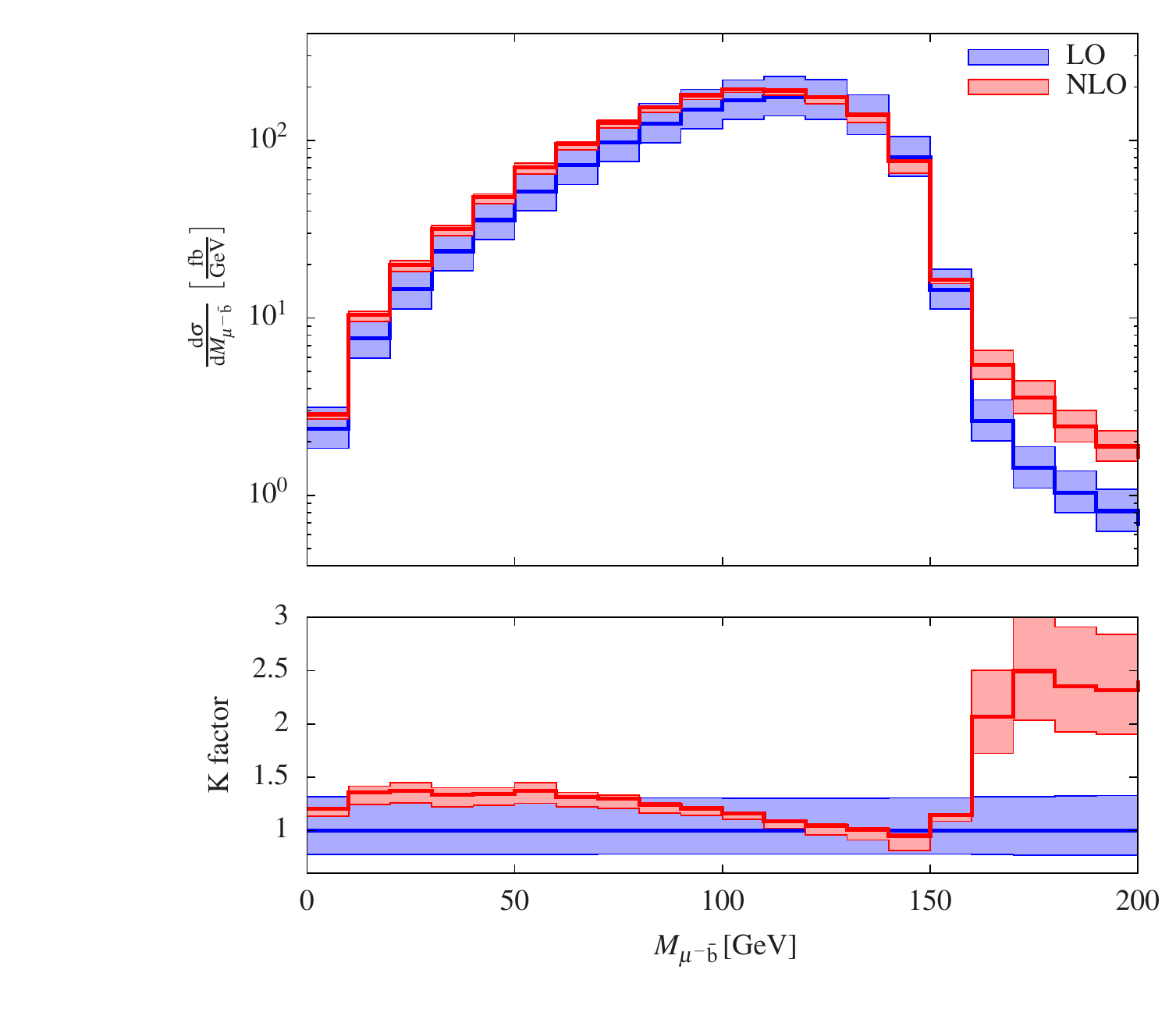}
                \label{plot:inv_mubx}
        \end{subfigure}        
        
        \vspace*{-3ex}
        \caption{\label{fig:inv_distributions}%
                Differential distributions at a centre-of-mass energy $\sqrt{s}=13\TeV$ at the LHC for $\Pp\Pp\to \mu^-\bar{\nu}_\mu\Pb\bar{\Pb} \Pj \Pj$: 
                \changed{
                \subref{plot:inv_antitop}~invariant mass of the leptonic top quark~(top left),
                \subref{plot:inv_tophad}~invariant mass of the hadronic top quark~(top right),
                \subref{plot:inv_jj}~invariant mass of the two hardest jets~(bottom left), and
                \subref{plot:inv_mubx}~invariant mass of the muon and anti-bottom quark~(bottom right).}
                The upper panels show the LO prediction as well as the NLO one.
                The lower panels display the ratio of the NLO and the LO predictions. The bands correspond to factor-2 scale
                variations as defined in Eq.~\eqref{eq:combination}.}
\end{figure}
In \reffi{fig:inv_distributions} \changed{four} invariant-mass
distributions are displayed.  Figure \ref{plot:inv_antitop} shows the
invariant mass of leptonically decaying top quark based on Monte Carlo
truth.  The well-known radiative tail below the top-quark resonance
(see, for instance, \citere{Denner:2012yc} for the off-shell
production of top-quark pairs that decay fully leptonically) is due to
final-state radiation that is not reconstructed with the decay
products of the top quark and thus not taken into account in the
definition of the top-quark invariant mass.  Hence, events that are
close to resonance at LO tend to be shifted below the resonance peak
when the extra real radiation is not soft or collinear to the bottom
quark.  Note that on the peak the corrections can reach about $-35\%$.
\changed{For the invariant mass of the hadronic top, this effect is
  enhanced owing to three QCD jets in the final state (two light jets
  and one bottom jet) that can radiate gluons instead of one bottom
  jet for the leptonic case.  As shown in \reffi{plot:inv_tophad},
  this results in negative NLO corrections of more than $100\%$ in some
  bins close to the resonance. A proper description of this
  distribution thus requires the inclusion of higher-order
  corrections, which is beyond the scope of this work.}

\changed{In \reffi{plot:inv_jj}, the distribution in the invariant
  mass of the two hardest jets (not necessarily the two jets that
  enter the definition of the hadronic top quark) is displayed.  At
  LO, the di-jet invariant-mass is restricted to the range $60{-}100
  \GeV$ owing to the invariant-mass cut on the di-jet system [see
  Eq.~\eqref{eq:mjjcuts}].  As explained above, this condition is
  relaxed at NLO due to the appearance of a third jet in the real
  emission.  Hence, these real contributions contribute in a wider
  part of phase space, while the LO and virtual contributions are
  restricted to the mass window $60{-}100\GeV$.  As a consequence the
  cancellation of enhanced IR-sensitive terms between virtual and real
  corrections is not happening locally.  In particular, this mechanism
  can lead to locally negative predictions at the kinematical
  boundaries \cite{Catani:1997xc} which
  is particularly apparent here in the bin between $80 \GeV$ and
  $120 \GeV$.}

The distribution in the invariant mass of the muon and the anti-bottom
quark is presented in \reffi{plot:inv_mubx}.  This observable has been
intensively investigated both in experiments \cite{Sirunyan:2017uhy}
and in theory \cite{Heinrich:2013qaa,Heinrich:2017bqp} as it has been
identified~\cite{Denner:2012yc} to be very sensitive to the value of
the top-quark mass.  Indeed, the sharp decrease at $154\GeV$
represents the transition from on-shell production of top quarks to a
region dominated by \changed{non-doubly-top-resonant
  contributions.\footnote{\changed{Above the threshold, the effect of
      non-doubly-top-resonant partonic channels amounts to
      $2{-}3\%$.}}} For on-shell top quark and W~boson, this invariant
mass is bounded at $M_{\mu^- \bar \Pb}^2 = M^2_{\rm t} - M^2_{\rm W}
\simeq \left(154 \GeV \right)^2$.  Below this limit, the NLO
corrections vary from $+35\%$ at $50\GeV$ to $-5\%$ at $150\GeV$ while
the $K$-factor increases up to $2.5$ above this limit.  This is simply
due to the fact that this region is dominated by non-resonant
contributions which receive large corrections as explained previously.
Hence the NLO corrections computed here are important and should have
an impact on the top-quark mass determination based on such an
observable \cite{Heinrich:2013qaa,Heinrich:2017bqp}.

\paragraph{Angular and rapidity distributions.}

\begin{figure}
        \setlength{\parskip}{-10pt}
        
        \begin{subfigure}{0.49\textwidth}
                \subcaption{}
                \includegraphics[width=\textwidth]{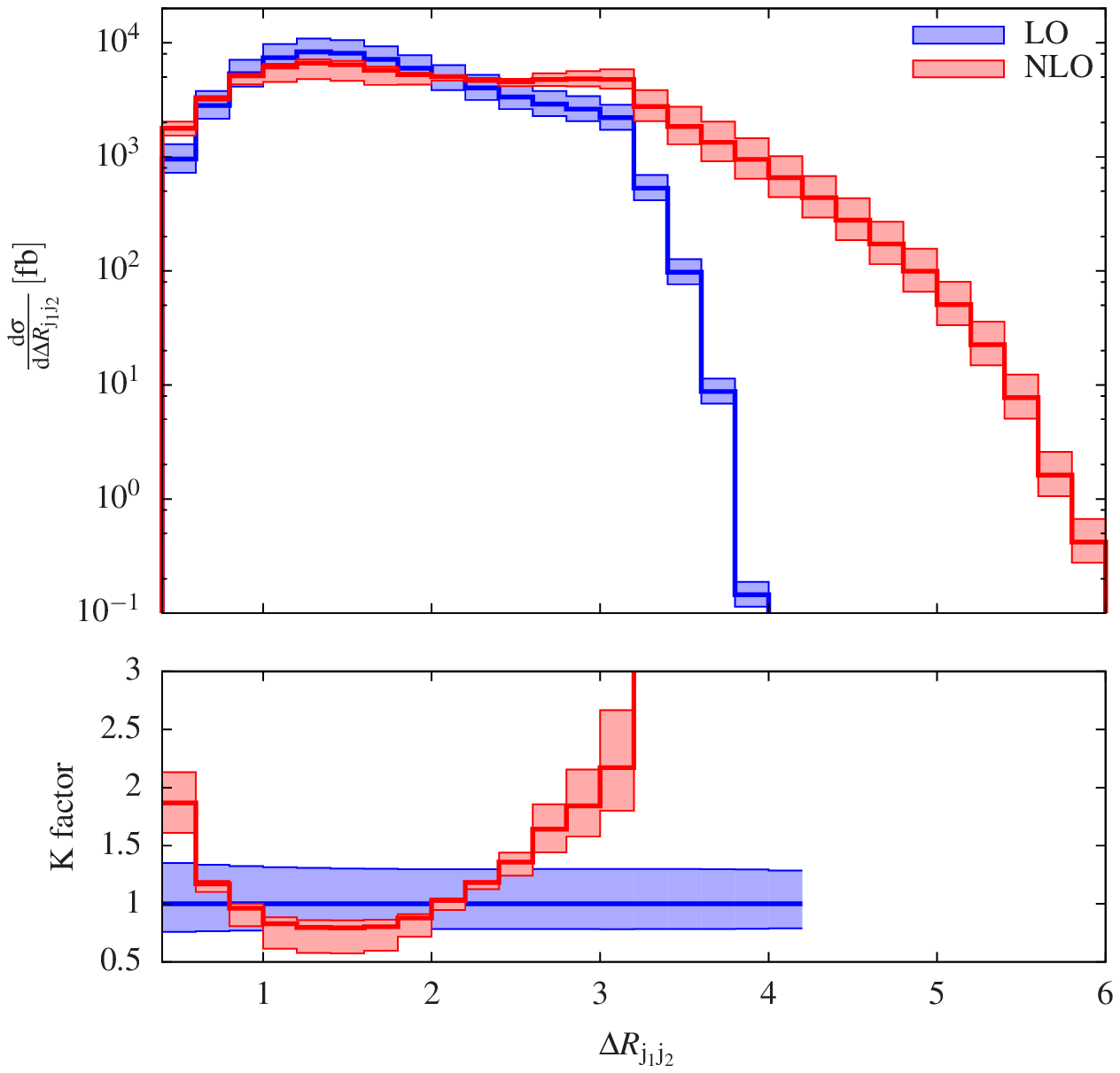}
                \label{plot:dist_j1j2}
        \end{subfigure}
        \hfill
        \begin{subfigure}{0.49\textwidth}
                \subcaption{}
                \includegraphics[width=\textwidth]{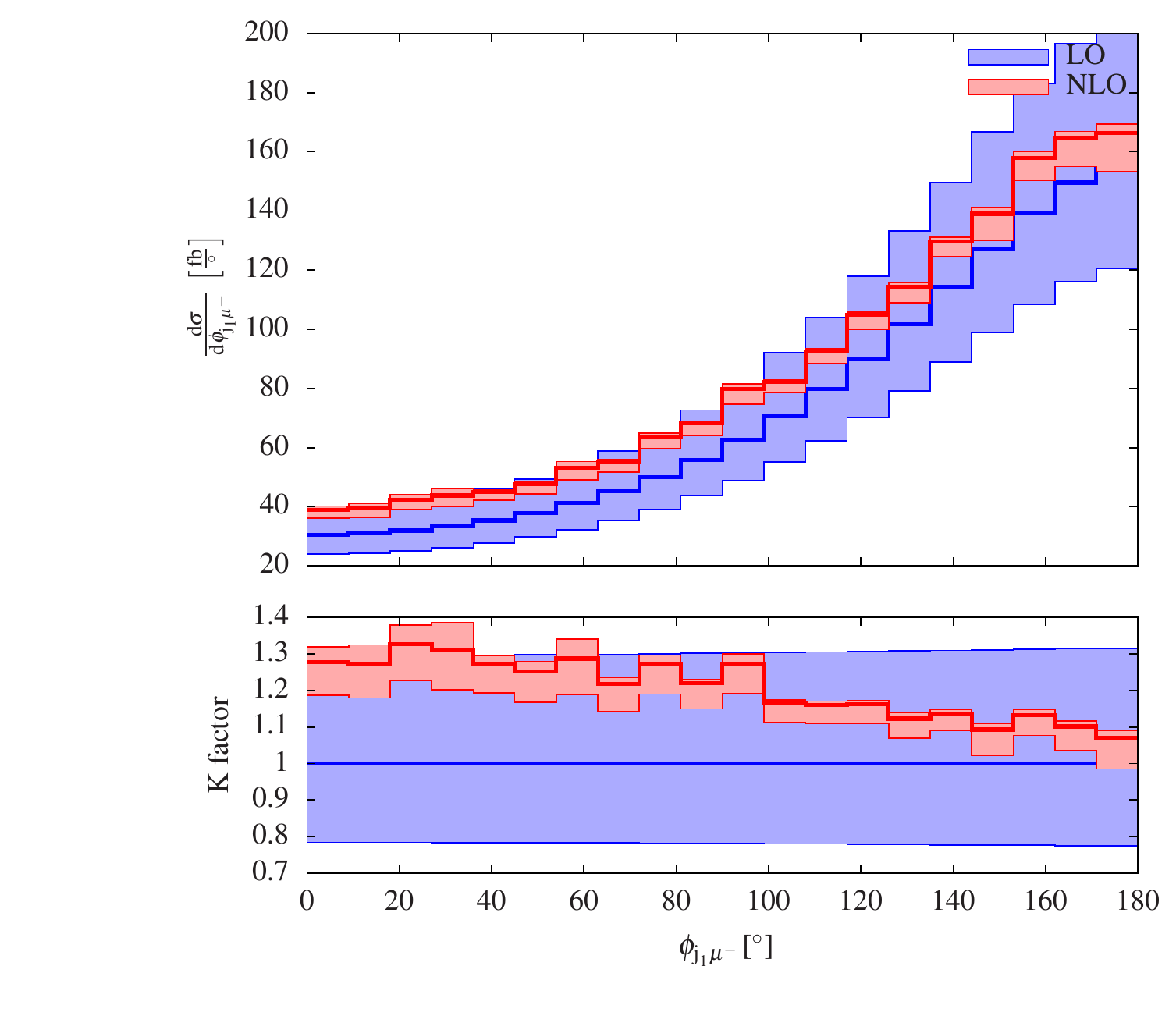}
                \label{plot:azi_j1mu}
        \end{subfigure}

        \begin{subfigure}{0.49\textwidth}
                \subcaption{}
                \includegraphics[width=\textwidth]{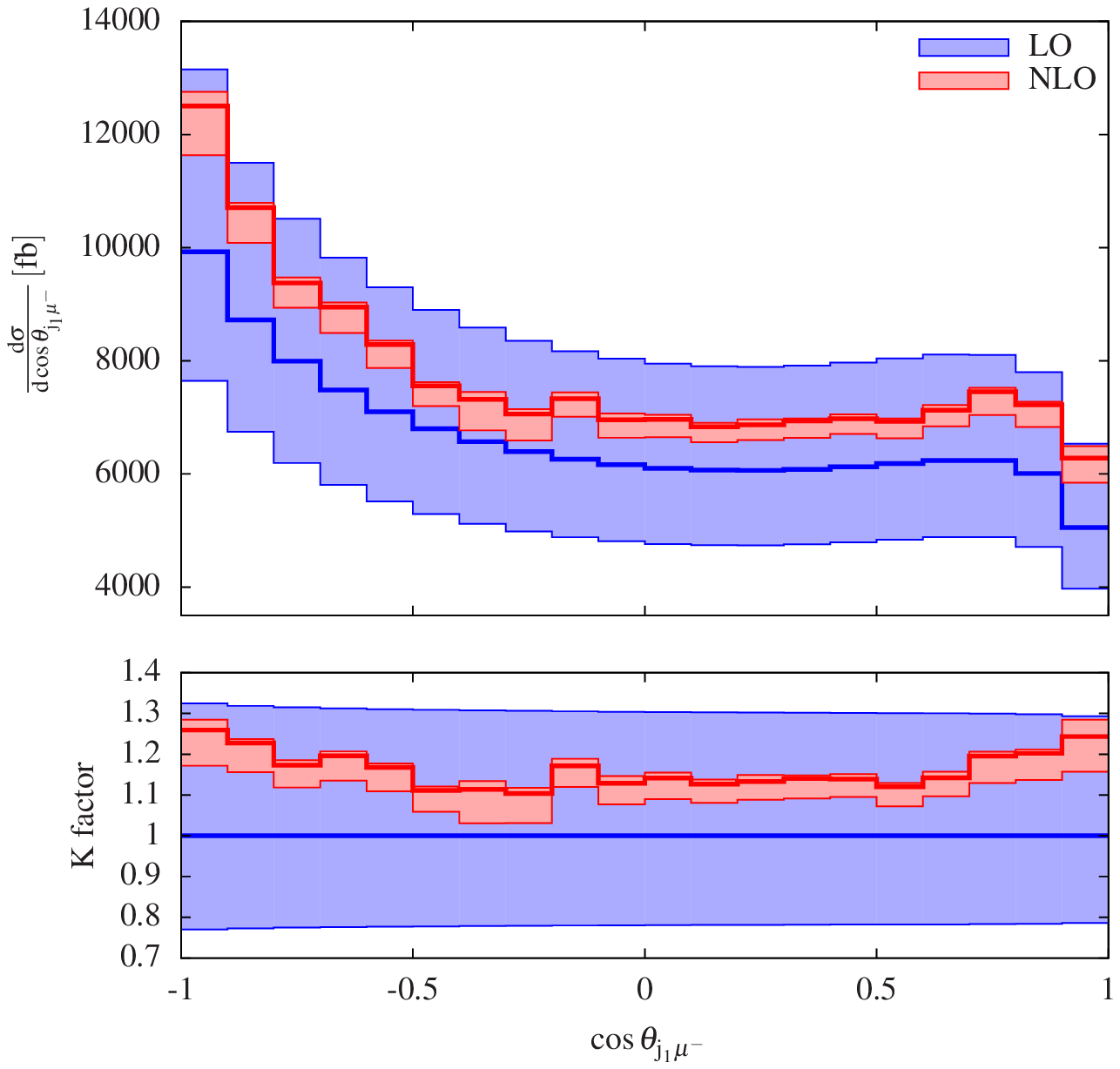}
                \label{plot:cos_j1mu} 
        \end{subfigure}
        \hfill
        \begin{subfigure}{0.49\textwidth}
                \subcaption{}
                \includegraphics[width=\textwidth]{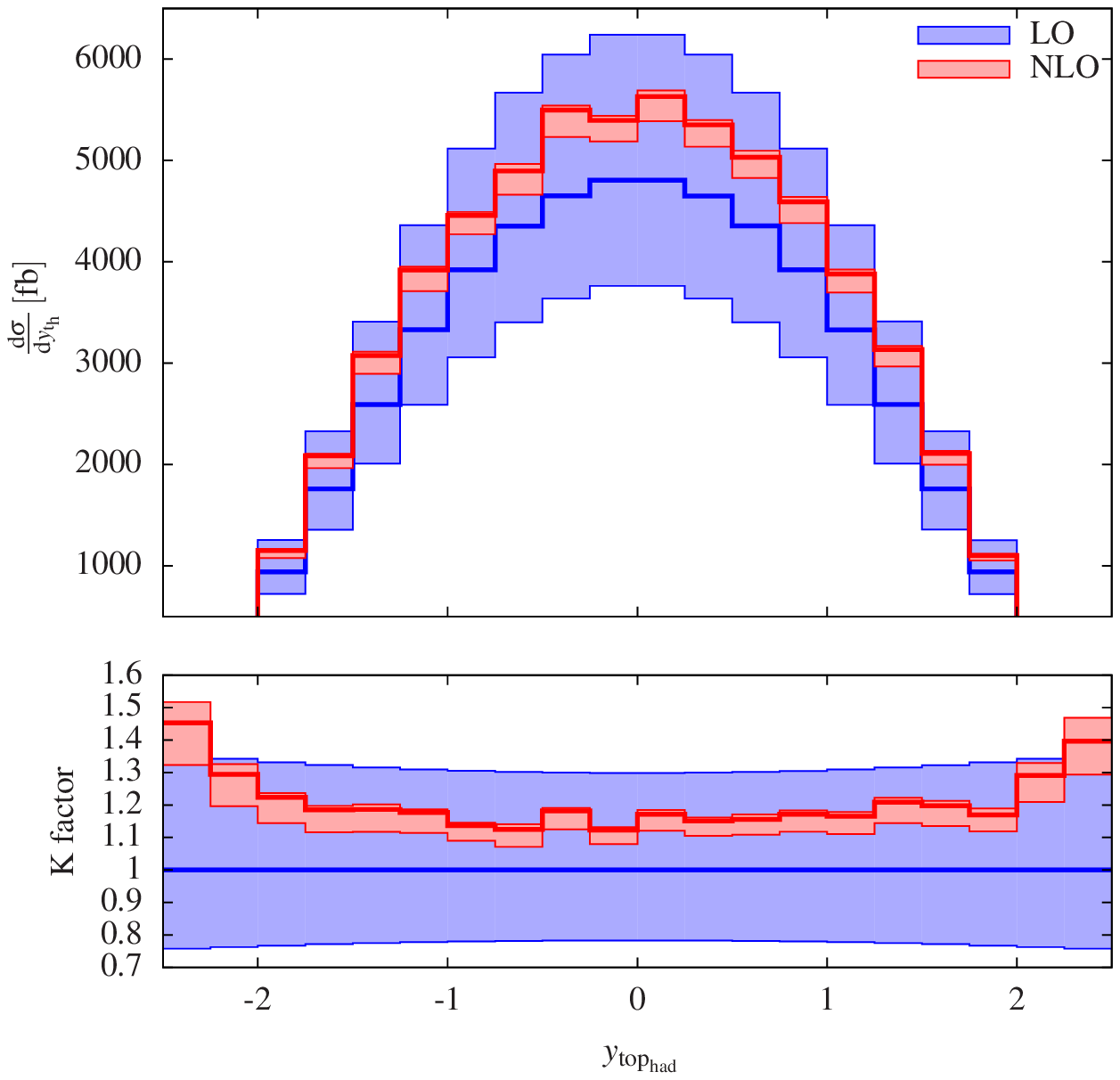}
                \label{plot:rap_top}
        \end{subfigure}
        
        \begin{subfigure}{0.49\textwidth}
                \subcaption{}
                \includegraphics[width=\textwidth]{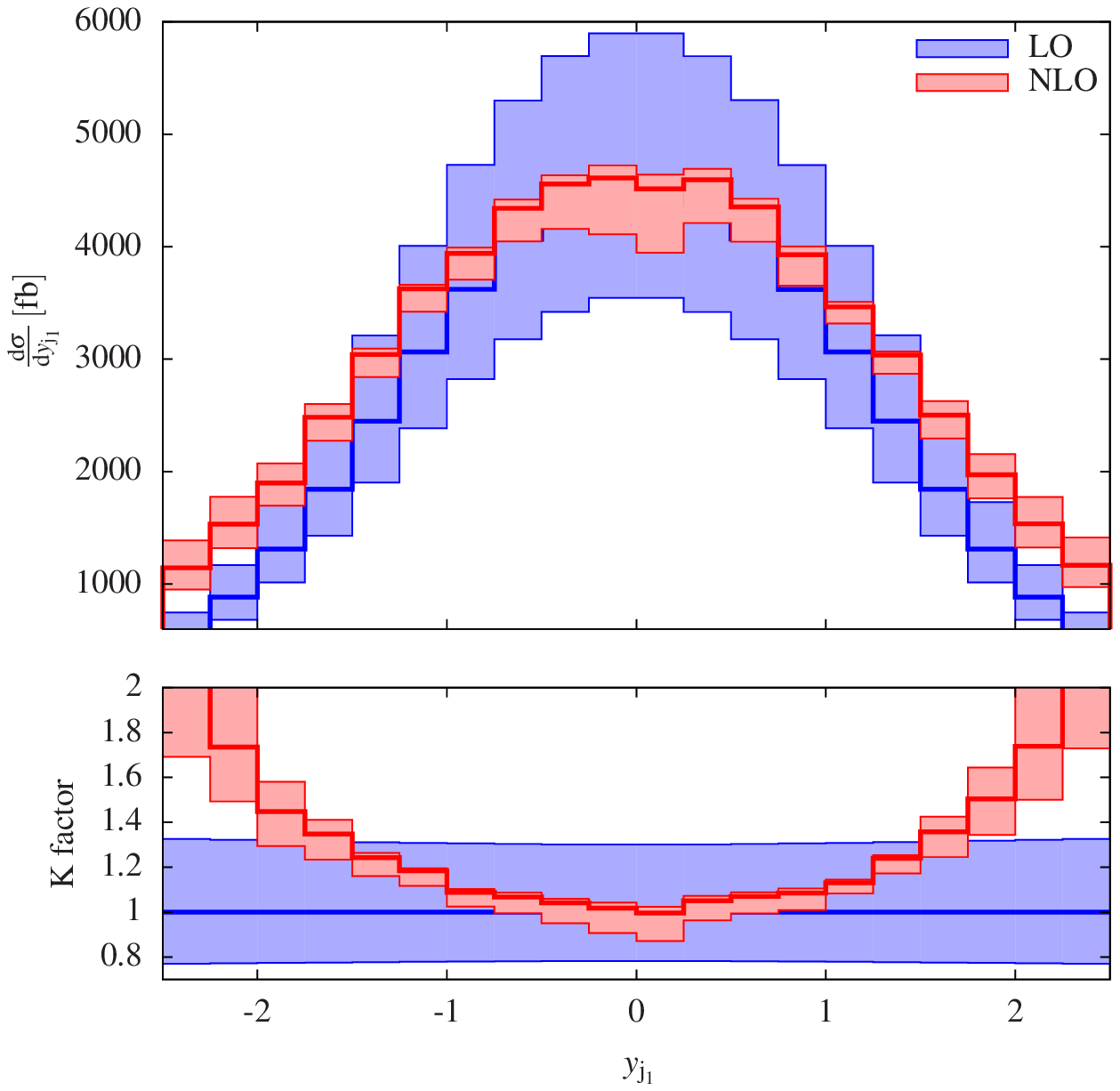}
                \label{plot:rap_j1}
        \end{subfigure}
        \hfill
        \begin{subfigure}{0.49\textwidth}
                \subcaption{}
                \includegraphics[width=\textwidth]{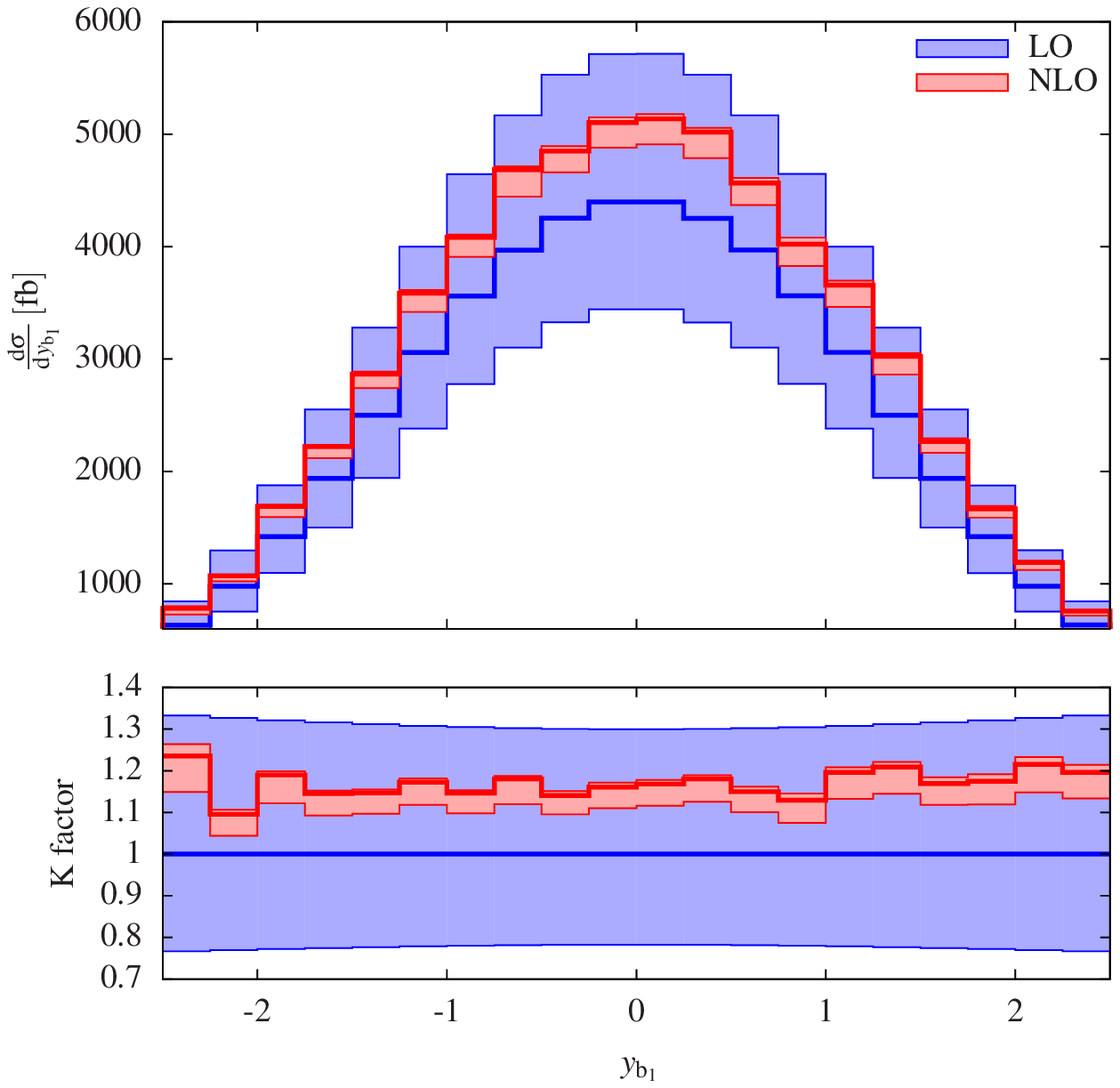}
                \label{plot:rap_b1}
        \end{subfigure}
        \vspace{-5ex}
                \caption{\label{fig:other_distributions}%
                Differential distributions at a centre-of-mass energy $\sqrt{s}=13\TeV$ at the LHC for $\Pp\Pp\to \mu^-\bar{\nu}_\mu\Pb\bar{\Pb} \Pj \Pj$: 
                \subref{plot:dist_j1j2}~rapidity--azimuthal-angle distance between the two hardest jets~(top left),
                \subref{plot:azi_j1mu}~azimuthal angle between the hardest jet and the muon~(top right),
                \subref{plot:cos_j1mu}~cosine of the angle between the hardest jet and the muon~(middle left).
                \subref{plot:rap_top}~rapidity of the reconstructed hadronic top quark~(middle right),
                \subref{plot:rap_j1}~rapidity of the hardest light jet~(bottom left), and
                \subref{plot:rap_b1}~rapidity of the hardest bottom jet~(bottom right).
                The upper panels show the LO prediction as well as the NLO one.
                The lower panels display the ratio of the NLO and the
                LO predictions. The bands correspond to factor-2 scale
                variations as defined in Eq.~\eqref{eq:combination}.}
\end{figure}

Finally, some angular and rapidity distributions are presented in
\reffi{fig:other_distributions}.  In \reffi{plot:dist_j1j2} we show
the distribution in the rapidity--azimuthal-angle distance between the
two hardest jets defined as in Eq.~\eqref{eq:distance}.  At LO, the
distribution is abruptly decreasing above $\pi$ and even going to zero
above 4.\changed{\footnote{\changed{Above $\Delta
      R_{\mathrm{j_1j_2}}=3.5$, the contribution of
      non-doubly-top-resonant partonic channels rises strongly to more
      than $5\%$.}}} Using again the approximation for small angles
gives $\Delta R_{\Pj\Pj,\rm max}^2 \sim {m_{\Pj \Pj,{\rm max}}^2} /
\ptsub{\Pj,\rm min}^2 = \left( 100 \right)^2 / \left( 25 \right)^2 =
4^2$.  Above this point, the LO contributions are forbidden due to the
event selection, but this is relaxed at NLO owing to the appearance of
real radiation as for the previously discussed distributions. As a
consequence the NLO contributions are dominating in this region.

Next, we present distributions in the azimuthal angle and in the
cosine of the angle between the hardest jet and the muon in
\reffis{plot:azi_j1mu} and \ref{plot:cos_j1mu}, respectively.  Both
distributions are smooth and relatively flat.  Therefore the NLO
corrections are pretty stable over the whole range.  They essentially
feature the normalisation present at the level of the fiducial cross
section.  In these distributions the reduction of the scale
uncertainty is particularly visible demonstrating the need for NLO
predictions.

We finish with rapidity distributions.  Figure \ref{plot:rap_top}
shows the rapidity distribution of the reconstructed hadronic top
quark, defined as for the transverse-momentum distribution shown in
\reffi{plot:pt_top_had}.  The corrections are smaller in the central
region where the two top quarks are mainly produced on shell.  In the
peripheral region, the non-resonant contributions come into play and
thus lead to larger corrections.  Finally, we present the rapidity
distributions of the hardest light and hardest bottom jet in
\reffis{plot:rap_j1} and \ref{plot:rap_b1}, respectively.  The NLO
corrections to the rapidity of the hardest jet vary stronger than those for
the hadronic top quark.  They reach $100\%$ at a rapidity of $\pm 2.5$
while they are close to zero in the central region.
In particular for $|y_{\rm j_1}|>2$ the LO and NLO uncertainty bands
do not overlap, and the NLO uncertainty band becomes larger in
this phase-space region.  
Large corrections, which can be attributed to real emission of jets from
the incoming partons, show up in this phase-space region such that the
accuracy of our prediction is effectively only of LO.  On the other
hand, the distribution in the rapidity of the hardest bottom jet does
not display significant shape distortions owing NLO corrections over
the whole kinematical range and merely inherits the correction factor
for the fiducial cross section.

\section{Conclusion}
\label{sec:conclusion}

The production of two top quarks gives rise to three different classes
of final states, depending on whether the two W~bosons and thus the
two top-quarks decay leptonically or hadronically.  So far, most of
the theoretical work has focused on the channel where the two top
quarks decay leptonically.  Nonetheless, the lepton+jets channel where
one top quark decays hadronically and the other leptonically possesses
some advantages over the fully leptonic channel.  First, it has a
larger cross section due to the hadronic branching ratio of the
W~boson and second, it allows for a better detection and
reconstruction of top quarks.  Indeed, in this case only one neutrino
in the final state
leads to missing transverse energy.  Hence, the process $\Pp\Pp\to
\mu^-\bar{\nu}_\mu\Pb\bar{\Pb} \Pj \Pj$ constitutes one of the key
channels for the study of the top-quark properties at the LHC, and
thus precise predictions for it are highly desirable.

So far, the full process was only known at LO.  At NLO, the best
available predictions included NLO corrections to the on-shell
production and LO decay of the top quarks.  \changed{ For the first
  time we have computed the NLO QCD corrections to the off-shell
  process $\Pp\Pp\to \mu^-\bar{\nu}_\mu\Pb\bar{\Pb} \Pj \Pj$ for all
  partonic channels that feature doubly-resonant top quarks.}  The
calculation features by definition off-shell and non-resonant effects.
These effects are becoming more and more relevant for run~II of the
LHC, where a large amount of data is collected at the increased energy
of $13\TeV$.  Hence the high-energy region where these effects are
more important will be accurately probed in the future, making such
computations very relevant.  In addition, in the present computation
the event selection applied to the final state mimics the one used by
the experimental collaborations in order to provide realistic
predictions.  \changed{We have focussed on the phase space relevant
  for top-pair production and omitted partonic channels that involve
  only one resonant top quark and one resonant W boson as well as
  bottom-quark-initiated and photon-induced contributions.  The
  corresponding leading-order contributions have been shown to be
  negligible in the phase space relevant for top-pair production.}

The results are different from those for leptonically decaying top
quarks, and the NLO corrections are particularly large in certain
phase-space regions.  These large corrections arise in high-energy
regions where contributions without two resonant top quarks are
important.  This is particularly explicit in the tails of the
transverse-momentum distributions.  Also, for observables that feature
a kinematic threshold, above this threshold the NLO corrections are
particularly large.  This happens for example for the distance between
the two hardest jets and for the invariant mass of the muon and
anti-bottom which is important for the top-quark mass determination.

The NLO corrections are very sensitive to the experimental event
selection.  In particular, the jet radius is a key parameter as it
ensures that the jets are separated in the resolved-topology event
selection.  Increasing or decreasing the jet radius would affect the
NLO corrections accordingly.  In addition, we have applied a cut on
the di-jet invariant mass ensuring that a jet pair originates most
probably from a W~boson and thus indirectly from a top quark.  This
cut, in particular, removes events at high transverse momentum where
the two jets originating from the W~boson are recombined in one jet
while the extra real radiation ensures the presence of two separated
jet.  In this way, mainly doubly-resonant top-quark contributions in
the resolved topology are selected at both LO and NLO.

On the technical side, the present NLO computation is non-trivial as
it possesses four coloured particles in addition to two leptons in the
final state.  Such a computation has been made possible by the use of
the matrix-element generator \recola in combination with the \collier
library as well as an efficient Monte Carlo program dubbed \mocanlo.

Finally, as the computation uses experimental event selection for the final states, this should allow the experimental collaborations to include these corrections in their forthcoming analysis.

\acknowledgments 

We are thankful to Jean-Nicolas Lang and Sandro Uccirati for
supporting \recola and Robert Feger concerning \mocanlo.  We thank
Mauro Chiesa for performing some checks of the calculation and Manfred
Kraus for useful discussions.  We acknowledge financial support by the
German Federal Ministry for Education and Research (BMBF) under
contract no.~05H15WWCA1 and the German Science Foundation (DFG) under
reference number DE 623/6-1.

\bibliographystyle{JHEPmod}
\bibliography{ttx_qcd} 

\end{document}

%% file: macros.tex
\newcommand{\Pl}{\ell}

\newcommand{\pb}{{\ensuremath\unskip\,\text{pb}}\xspace}
\def\reffi#1{\mbox{Figure~\ref{#1}}}
\def\reffis#1{\mbox{Figures~\ref{#1}}}
\def\refta#1{\mbox{Table~\ref{#1}}}

\def\refse#1{\mbox{Section~\ref{#1}}}
\def\refses#1{\mbox{Sections~\ref{#1}}}

\def\citere#1{\mbox{Ref.~\cite{#1}}}
\def\citeres#1{\mbox{Refs.~\cite{#1}}}

\newcommand{\ie}{\emph{i.e.}\ }

\def\be{\begin{equation}}
\def\ee{\end{equation}}

\newcommand{\Pj}{\ensuremath{\text{j}}\xspace}
\newcommand{\Pp}{\ensuremath{\text{p}}\xspace}
\newcommand{\Pe}{\ensuremath{\text{e}}\xspace}
\newcommand{\Pb}{\ensuremath{\text{b}}\xspace}

\newcommand{\Pt}{\ensuremath{\text{t}}\xspace}
\newcommand{\Pu}{\ensuremath{\text{u}}\xspace}
\newcommand{\Pd}{\ensuremath{\text{d}}\xspace}
\newcommand{\Ps}{\ensuremath{\text{s}}\xspace}
\newcommand{\Pc}{\ensuremath{\text{c}}\xspace}
\newcommand{\Pg}{\ensuremath{\text{g}}\xspace}

\newcommand{\PW}{\ensuremath{\text{W}}\xspace}
\newcommand{\PZ}{\ensuremath{\text{Z}}\xspace}

\newcommand{\Mt}{\ensuremath{m_\Pt}\xspace}

\newcommand{\MWOS}{\ensuremath{M_\PW^\text{OS}}\xspace}
\newcommand{\MW}{\ensuremath{M_\PW}\xspace}
\newcommand{\MZOS}{\ensuremath{M_\PZ^\text{OS}}\xspace}
\newcommand{\MZ}{\ensuremath{M_\PZ}\xspace}

\newcommand{\Gt}{\ensuremath{\Gamma_\Pt}\xspace}

\newcommand{\GZOS}{\ensuremath{\Gamma_\PZ^\text{OS}}\xspace}

\newcommand{\GWOS}{\ensuremath{\Gamma_\PW^\text{OS}}\xspace}

\newcommand{\MVOS}{\ensuremath{M_V^\text{OS}}\xspace}%
\newcommand{\GVOS}{\ensuremath{\Gamma_V^\text{OS}}\xspace}%

\newcommand{\GeV}{\ensuremath{\,\text{GeV}}\xspace}
\newcommand{\TeV}{\ensuremath{\,\text{TeV}}\xspace}

\newcommand{\alphas}{\ensuremath{\alpha_\text{s}}\xspace}
\newcommand{\order}[1]{\ensuremath{\mathcal{O}{\left(#1\right)}}\xspace}

\newcommand{\GF}{\ensuremath{G_\mu}}

\newcommand{\ptsub}[1]{\ensuremath{p_{\text{T},#1}}\xspace}

\renewcommand{\Re}{\mathop{\mathrm{Re}}\nolimits}

\newcommand{\Etbar}{\overline{E_\mathrm{T}}}

\newcommand{\recola}{{\sc Recola}\xspace}
\newcommand{\mocanlo}{{\sc MoCaNLO}\xspace}
\newcommand{\collier}{{\sc Collier}\xspace}

\newcolumntype{.}{D{.}{.}{-1}}
\newcolumntype{d}[1]{D{.}{.}{#1}}

\colorlet{tableoverheadcolor}{gray!37.5}
\colorlet{tableheadcolor}{gray!25}
\colorlet{tablerowcolor}{gray!12.5}

\newlength{\width}
\newlength{\height}

\def\draftdate{\relax}
\def\mda{\relax}
\def\mua{\relax}
\def\mla{\relax}
\def\draft{
\def\thtystars{******************************}
\def\sixtystars{\thtystars\thtystars}
\typeout{}
\typeout{\sixtystars**}
\typeout{* Draft mode!
         For final version remove \protect\draft\space in source file *}
\typeout{\sixtystars**}
\typeout{}
\def\draftdate{\today}
\def\mua{\marginpar[\boldmath\hfil$\uparrow$]%
                   {\boldmath$\uparrow$\hfil}\color{black}%
                    \typeout{marginpar: $\uparrow$}\ignorespaces}
\def\mda{\color{red}\marginpar[\boldmath\hfil$\downarrow$]%
                   {\boldmath$\downarrow$\hfil}%
                    \typeout{marginpar: $\downarrow$}\ignorespaces}
\def\mla{\marginpar[\boldmath\hfil$\rightarrow$]%
                   {\boldmath$\leftarrow $\hfil}%
                    \typeout{marginpar: $\leftrightarrow$}\ignorespaces}
\def\Mua{\marginpar[\boldmath\hfil$\Uparrow$]%
                   {\boldmath$\Uparrow$\hfil}\color{black}%
                    \typeout{marginpar: $\uparrow$}\ignorespaces}
\def\Mda{\color{red}\marginpar[\boldmath\hfil$\Downarrow$]%
                   {\boldmath$\Downarrow$\hfil}%
                    \typeout{marginpar: $\downarrow$}\ignorespaces}
\def\Mla{\marginpar[\boldmath\hfil\textcolor{red}{$\Rightarrow$}]%
                   {\boldmath\textcolor{red}{$\Leftarrow $}\hfil}%
                    \typeout{marginpar: $\leftrightarrow$}\ignorespaces}
\overfullrule 5pt
\oddsidemargin 15mm
\marginparwidth 29mm
}
